\documentclass[journal]{IEEEtran}
\usepackage[T1]{fontenc}
\usepackage{newtxtext}
\usepackage{graphicx}    % 插图
\usepackage{bm}          % bold mathematical entries in tables
\usepackage{epstopdf} % 自动转换EPS为PDF

\usepackage{capt-of}     % captions for the compact Table VII--VIII layout
\ifCLASSINFOpdf

\graphicspath{../png/}
\DeclareGraphicsExtensions{.png}
\else
\fi
\usepackage{diagbox}
\usepackage{stfloats}
\usepackage{placeins}
\usepackage{booktabs}
\usepackage{cite}
\usepackage{url}
\usepackage[hypertexnames=false]{hyperref}
\hypersetup{hidelinks}
\usepackage{array}
\usepackage{multirow} % 确保导言区引入该包
\usepackage{graphics}
\usepackage{amssymb,amsmath}
\usepackage{makecell}
\usepackage{color}
\usepackage{changepage}
\usepackage{colortbl}
\usepackage{algorithm}
\usepackage{algorithmic}
\usepackage[table]{xcolor}
\definecolor{red}{rgb}{0,0,0}
\definecolor{revisionRtwo}{rgb}{0,0,0}
\definecolor{magenta}{rgb}{0,0,0}
\usepackage[normalem]{ulem}
\usepackage{amsthm}

\theoremstyle{definition}
\newtheorem{definition}{Definition}

\theoremstyle{remark}

\newcolumntype{I}{!{\vrule width 3pt}}
\newlength\savedwidth

\newlength\savewidth

\newcommand{\meanstd}[2]{\ensuremath{#1\,{\scriptstyle\pm\,#2}}}
\newcommand{\meanstdbf}[2]{\ensuremath{\bm{#1}\,{\scriptstyle\pm\,#2}}}

\begin{document}
% Keep the two text columns aligned at the page bottom.
\flushbottom
\bstctlcite{BSTcontrol}
\title{
CoSyDelay: Cognitive Symbolic Regression for Traffic Delay Estimation 
}

% \title{Automated Surrogate-based Optimization Algorithm with Large Language Models for Traffic Signal Optimization

% Surrogate-Based Traffic Signal Optimization with Large Language Models
% }

\author{Lindong Xie,
        Xiexin Zou,
        Yang Zhang,
        Beiyu Song,
        Zeren Xing, and
        Edward~Chung*
       % <-this % stops a space

\thanks{This work was supported by the Hong Kong Research Grants Council - General Research Fund (Grant No. 15224423) and the Innovation and Technology Commission - Mainland-Hong Kong Joint Funding Scheme (Grant No. MHP/038/23). \textit{(Corresponding author: Edward Chung.)}}
        
\thanks{L.~Xie, X. Zou, Y. Zhang, B. Song, Z.~Xing, and E.~Chung are with The Hong Kong Polytechnic University, Hong Kong, China (email: lindong.xie@connect.polyu.hk; edward.cs.chung@polyu.edu.hk).}% <-this % stops a space
}

\markboth{IEEE}
{CoSyDelay: Cognitive Symbolic Regression for Traffic Delay Estimation}
\maketitle
\newcommand{\myfootnote}[2]{
    \renewcommand{\thefootnote}{}
    \footnotetext{\footnotesize#1}
    \renewcommand{\thefootnote}{\arabic{footnote}}
}

\begin{abstract}

Trustworthy estimation of traffic delays at signalized intersections is critical for modern transportation systems, as it supports proactive traffic control for congestion mitigation. Existing analytical methods often rely on idealized assumptions that limit their accuracy under complex traffic conditions, whereas data-driven methods may achieve strong estimation performance but usually lack transparent structures, which reduces confidence in their practical application. To address these limitations, we propose \textbf{CoSyDelay}, a \uline{co}gnitive \uline{sy}mbolic regression (SR) framework for automatically deriving explicit and physically consistent empirical delay estimation expressions (DEEs). Specifically, CoSyDelay decomposes intersection-level delay estimation into turning-movement-level subproblems, thereby reducing the complexity of the symbolic search. During DEE learning, CoSyDelay constructs domain-informed symbolic inputs from traffic variables. Large language models (LLMs) are employed as symbolic search operators to iteratively generate and refine candidate expression structures, whose movement-specific coefficients are subsequently calibrated using approach-level delay labels. A composite fitness function evaluates the learned DEEs jointly in terms of estimation accuracy and compliance with specified traffic-delay principles. Experimental results on multiple signalized intersections demonstrate that the resulting expressions achieve competitive estimation performance while maintaining physical consistency, thereby supporting trustworthy delay estimation. \textcolor{red}{Source code and datasets are available at \url{https://github.com/ForrestXie9/CoSyDelay}.}

\end{abstract}


\IEEEpeerreviewmaketitle

\section{Introduction}
In urban road networks, signalized intersections are critical nodes that strongly influence traffic efficiency and congestion patterns \cite{yue2021root,hong2022traffic}. Traffic delay, commonly defined as the additional travel time experienced by vehicles relative to free-flow conditions \cite{akcelik1981traffic}, is a key indicator of intersection performance. It has direct economic implications, including increased fuel consumption and travel time costs, while remaining intuitive to both technical and non-technical stakeholders \cite{hurdle1984signalized,cheng2016review}. As direct measurement of delay at signalized intersections is difficult, trustworthy delay estimation methods are essential \cite{su2009delay}. In particular, such methods should provide accurate estimates while maintaining explicit, physically consistent model structures for traffic engineering use.

\textcolor{red}{Existing methods for traffic delay estimation are broadly categorized as analytical or data-driven. Analytical models, including Webster's model \cite{webster1958traffic}, Ak\c{c}elik's model \cite{akcelik1981traffic}, and the Highway Capacity Manual procedure \cite{manual2010hcm2010}, estimate delay by substituting traffic demand, signal timing, and intersection geometry into manually derived delay equations. Their explicit structures and modest computational requirements have led to their widespread use for routine assessment of signalized-intersection performance \cite{cheng2016review}. However, simplifying assumptions and fixed functional forms may reduce their accuracy under complex and time-varying traffic conditions \cite{dion2004comparison,teply1989accuracy}. Data-driven methods learn delay relationships from observations using machine learning or deep learning \cite{murat2006comparison,subramaniyan2022hybrid}. They can capture complex nonlinear interactions and achieve high estimation accuracy \cite{almukhalfi2024traffic}. Such models often encode these relationships implicitly, making their estimation mechanisms difficult to inspect and offering no guarantee of compliance with traffic-delay principles \cite{makke2024interpretable,ciatto2024symbolic}.}

\textcolor{red}{Symbolic regression (SR) learns explicit mathematical expressions from data, making the learned relationships directly inspectable \cite{zhang2023deep,makke2024interpretable}. To guide SR toward expressions consistent with known relationships, knowledge-guided SR incorporates prior knowledge into candidate generation or evaluation through grammar restrictions \cite{crochepierre2022reinforcement}, dimensional constraints \cite{tenachi2023deep}, shape constraints \cite{kronberger2022shape}, or formal verification \cite{bladek2023counterexample}. Two gaps remain when applying knowledge-guided SR to signalized-intersection delay estimation. First, a unified intersection-level delay expression must combine traffic and signal variables across multiple turning movements and phases. This creates a high-dimensional search space in which existing symbolic-search methods struggle to identify accurate expressions \cite{virgolin2022symbolic,makke2024interpretable}. Second, existing knowledge-guided SR methods generally do not represent traffic-delay-specific principles in their search and evaluation procedures. An expression can therefore fit observed delays while violating required delay behavior \cite{bartlett2023exhaustive}.}

\textcolor{red}{To address these gaps, we propose \textbf{CoSyDelay}, a cognitive symbolic regression framework that automates the iterative process of proposing, calibrating, checking, and revising traffic-delay formulas. CoSyDelay addresses the first gap by decomposing an intersection into turning-movement-level SR subproblems, avoiding a single search over all movement and phase variables. For each subproblem, it transforms the raw inputs into the flow ratio, green ratio, and cycle length. These variables are established traffic-theory inputs for delay estimation \cite{webster1958traffic,akcelik1981traffic,manual2010hcm2010,chandler2013signalized}. Subsequently, CoSyDelay employs LLM-guided SR to propose plausible candidate expression structures from these variables \cite{merler2024incontext,shojaee2024llm,zhang2026llm}. The movement-specific coefficients of each structure are then jointly calibrated from approach-level delay labels. CoSyDelay addresses the second gap by introducing a customized external evaluator. It evaluates each candidate in two respects: estimation accuracy and compliance with traffic-delay principles (e.g., non-negativity, monotonicity, and boundary behavior). The two evaluation scores are combined in the composite fitness function to select candidates for the next iteration, while estimation-accuracy results and principle-level diagnostics jointly guide LLM revisions. After the iterative search terminates, the resulting movement-level delay estimates are aggregated using traffic-volume weights to estimate approach- and intersection-level delays \cite{manual2010hcm2010,fhwa2008traffic}.}

The main contributions of this work are as follows:

\begin{itemize}
    \item We propose \textbf{CoSyDelay}, a novel SR--based framework that learns physically consistent empirical delay expressions at the turning-movement level and enables multi-granularity delay estimation at the movement, approach, and intersection levels.

    \item We reduce symbolic search difficulty through turning-movement-level decomposition, domain-informed SR inputs, and LLM-guided symbolic search. We further design a customized fitness function that jointly considers estimation accuracy and traffic-delay principles.

    \item We conduct extensive experiments on diverse signalized intersections. The results show that \textbf{CoSyDelay} achieves competitive performance against conventional methods while producing physically consistent empirical delay expressions.

\end{itemize}

The remainder of this paper is organized as follows.
Section \uppercase\expandafter{\romannumeral2} provides the necessary preliminaries.
Section \uppercase\expandafter{\romannumeral3} reviews the relevant literature.
Section \uppercase\expandafter{\romannumeral4} introduces the proposed algorithm.
Section \uppercase\expandafter{\romannumeral5} presents the numerical experiments and empirical evaluation results.
Finally, Section \uppercase\expandafter{\romannumeral6} concludes the paper and outlines potential directions for future research.

\section{Preliminaries}
\subsection{Traffic Delay Estimation Problem}

\begin{figure}[!htbp]
        \centering
\includegraphics[height=1.55 in,width=3.35in]{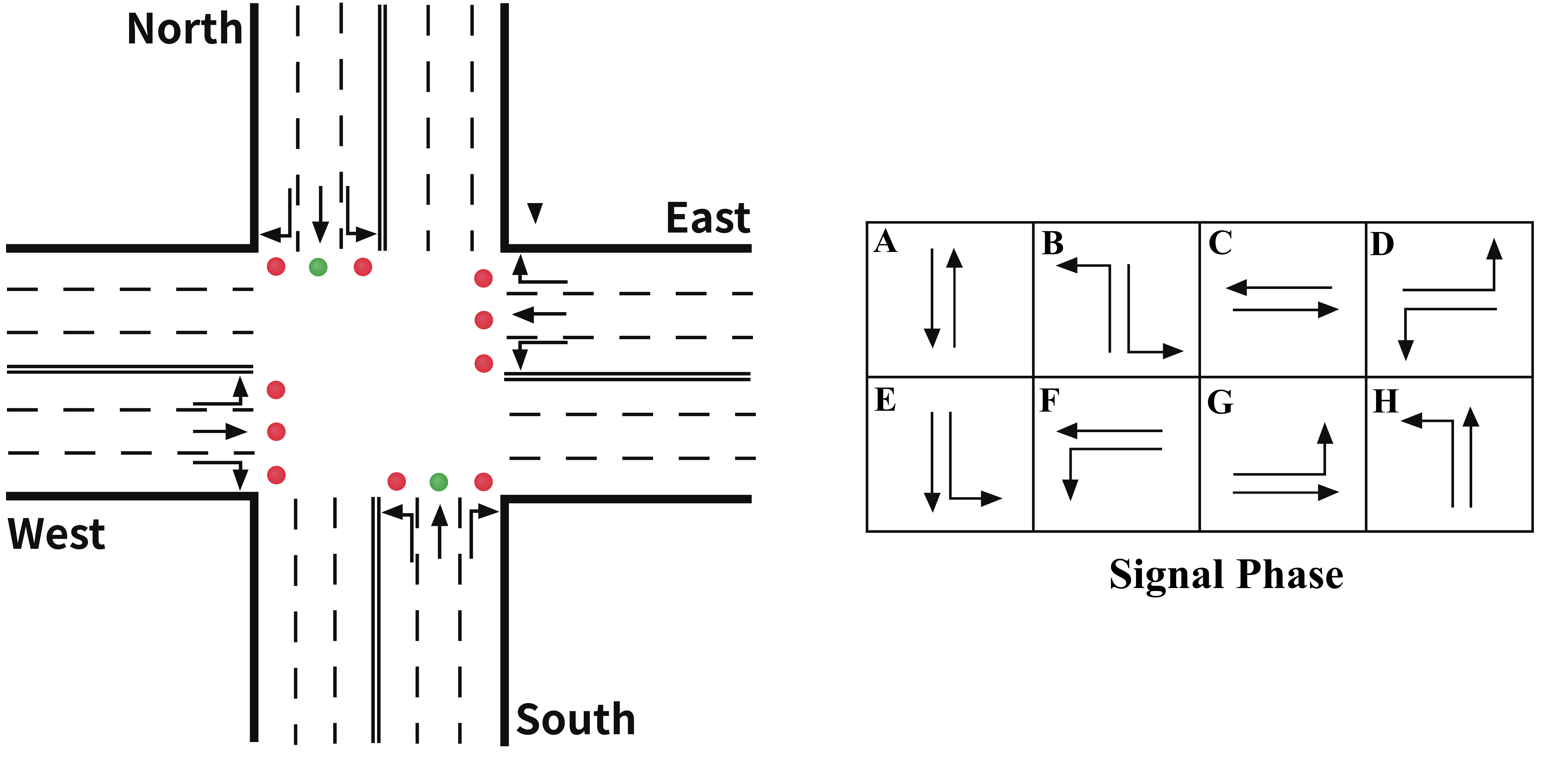}
\caption{Illustration of a four-approach intersection (South, West, North, and East) with eight traffic signal phasing settings (Phase A to Phase H). In this example, Phase A is activated to serve the corresponding traffic movements.}
        \label{fig:Layout_Inter}
\end{figure}

We begin with some basic definitions used in the paper.

\begin{definition}[Traffic turning movement]
A \emph{turning movement} is a directional vehicle flow from an entry approach to an exit approach at an intersection. Turning movements are commonly classified as \emph{through}, \emph{left-turn}, and \emph{right-turn} according to the relative orientation of the entry and exit approaches. As shown in Fig.~\ref{fig:Layout_Inter}, the illustrative intersection contains 12 distinct turning movements.
\end{definition}

\begin{definition}[Signal phase]
A \emph{signal phase} is a time interval during which a predefined set of non-conflicting turning movements is granted the right-of-way. Each phase consists of \emph{green}, \emph{yellow}, and \emph{all-red} intervals, which together determine the service time allocated to the associated movements. The intersection in Fig.~\ref{fig:Layout_Inter} has eight phases, each serving its assigned movements simultaneously.
\end{definition}

\begin{definition}[Signal cycle]
A \emph{signal cycle} is a complete, periodically repeated sequence of signal phases; its duration is the cycle length.
\end{definition}

\begin{definition}[Saturation flow rate and flow ratio]
The \emph{saturation flow rate} is the maximum discharge rate that can be achieved by a lane group or turning movement under continuous green and ideal operating conditions. The \emph{flow ratio} is the ratio of traffic volume to the corresponding saturation flow rate, reflecting the demand-to-capacity relationship of a given movement.
\end{definition}

\begin{definition}[Traffic delay]
Traffic delay is measured in this study as the time loss accumulated within the approach detection area relative to a reference travel speed. In this paper, we focus on movement-level delay estimation, which can be written as
\begin{equation}
\left\{ \mathbf{x}_1, \dots, \mathbf{x}_n \right\} 
\overset{f(\cdot)}{\longmapsto} 
\left\{\hat d_1, \dots, \hat d_n \right\},
\label{eq:Delay_Problem}
\end{equation}
where \(\mathbf{x}_i=(q_i,g_i^{\mathrm{eff}},C_i)^\top\in\mathbb{R}^{3}\) denotes the feature vector of the \(i\)-th turning movement, containing traffic volume, effective green time, and cycle length. \(\hat d_i\) denotes the estimated delay of movement \(i\). The function \(f(\cdot)\) maps the movement features \(\mathbf{x}_i\) to the corresponding estimate \(\hat d_i\).
\end{definition}

\subsection{Symbolic Regression}

SR seeks a mathematical expression that best captures the relationship between input and output data \cite{wu2023discovering}. Given a dataset \( \mathcal{DB} = \{ (\mathbf{x}_{i}, z_{i}) \}_{i=1}^N \), where \( \mathbf{x}_{i} \in \mathbb{R}^{d_x} \) is the input feature vector and \( z_{i} \in \mathbb{R} \) is the target value, the goal is to identify an optimal symbolic model \( f^* \in \mathcal{F} \) from a predefined candidate space \( \mathcal{F} \), as described in \cite{makke2024interpretable}:
\begin{equation}
f^* = \arg \min_{f \in \mathcal{F}} \mathcal{R}(f),
\label{eq:SR_Problem}
\end{equation}
where \( f^* \) is the optimal symbolic function. The candidate space \( \mathcal{F} \) consists of variables, free constants, basic arithmetic operators, and mathematical functions. The empirical loss \( \mathcal{R}(f) = \frac{1}{N} \sum_{i=1}^{N} \ell(f(\mathbf{x}_i), z_i) \) measures the average discrepancy between the predicted and observed outputs over the dataset \( \mathcal{DB} \).

\section{Literature Review}

\subsection{Traffic Delay Estimation Methods}
\subsubsection{Analytical Model for Delay Estimation}
Analytical models for traffic delay estimation predict vehicle delays at intersections using mathematical formulations grounded in traffic theory. For instance, Webster's delay model \cite{webster1958traffic} estimates average control delay by relating the signal cycle length, effective green time, and traffic arrival rate, assuming uniform vehicle arrivals and computing delay as the time vehicles spend waiting at red lights. Building on this, Miller's model \cite{miller1963settings} incorporated the effects of non-uniform and fluctuating arrivals, with the aim of improving applicability under realistic traffic conditions. Ak\c{c}elik's model \cite{akcelik1981traffic} extends signal-delay estimation to oversaturated and time-dependent queueing conditions, improving its applicability under high demand. Similarly, the Highway Capacity Manual (HCM) control delay model \cite{manual2010hcm2010} decomposes total delay into uniform delay due to regular signal stops, random delay arising from stochastic variations in vehicle arrivals, and overflow delay caused by queues exceeding the green time. Each component is calculated using traffic volume, saturation flow, and signal timing, enabling a detailed representation of intersection performance. Although these methods are grounded in traffic theory and offer strong interpretability, their simplified model structures often struggle to provide accurate predictions when faced with complex and dynamically changing traffic conditions \cite{ranpura2024estimation}.

\subsubsection{Data-Driven Models for Delay Estimation}
\textcolor{red}{Data-driven methods are widely used in traffic forecasting, where historical observations across road networks are used to predict future flow, speed, or demand. Recent graph-based approaches model complex spatiotemporal dependencies through space-specific and multi-view graph learning \cite{yang2024ssgcrtn,yang2025mstdfgrn,yang2025pstcgcn}, temporal-identity encoding and multiscale interaction \cite{yang2025tiidgcn,yang2025sdsinet,yang2025mtegcrn}, and decoupled dynamic graph learning \cite{yang2025gdgcrn,yang2026dmfgcrn}. Related data-driven studies also use multi-representation road-network features for trajectory recovery \cite{zhou2025mredtrajrec} and station-aware graph attention for cellular traffic prediction \cite{yang2026sgaseq}. These studies demonstrate the flexibility of graph-based representation learning, but they do not derive explicit traffic-delay equations. Data-driven techniques have also been applied to signalized-intersection delay estimation. Murat \cite{murat2006comparison} and Preethi et al.~\cite{preethi2016modelling} develop artificial neural network (ANN) models using traffic demand, signal timing, geometric characteristics, and operating conditions. Bagdatli and Dokuz~\cite{bagdatli2021vehicle} and Ranpura et al.~\cite{ranpura2024estimation} investigate machine-learning methods for signalized-intersection delay estimation. Korkmaz et al.~\cite{korkmaz2017delay,korkmaz2020comparison} adopt a semi-empirical approach, using heuristic algorithms to calibrate predefined linear, exponential, and quadratic delay functions based on the green ratio and degree of saturation.}

\textcolor{red}{Existing data-driven approaches exhibit a trade-off between modeling flexibility and structural transparency. Graph-based forecasting and machine-learning delay models can learn complex relationships but encode them implicitly. Semi-empirical delay models yield explicit equations but confine learning to predefined functional forms. Consequently, methods that discover explicit delay-equation structures directly from data remain limited.}

\subsection{Knowledge-Guided Model Discovery}

\subsubsection{Symbolic Regression and Knowledge-Guided Search}
\textcolor{red}{SR discovers explicit mathematical expressions directly from data, making the learned relationships inspectable \cite{dong2025recent}. Traditional SR searches expression structures through genetic programming, while deep SR uses learned generative models \cite{petersen2019deep}. In both cases, the search space grows rapidly with the available variables and operators \cite{virgolin2022symbolic,makke2024interpretable}. Accuracy and expression complexity alone do not ensure consistency with known domain relationships. Knowledge-guided SR addresses this problem through grammar rules \cite{crochepierre2022reinforcement}, dimensional constraints \cite{tenachi2023deep}, shape constraints \cite{kronberger2022shape}, or formal verification \cite{bladek2023counterexample}. Related structured-learning studies encode knowledge in model representations through ontology-enhanced subgraphs, temporal graph--text alignment, and relation-guided multimodal fusion \cite{wang2026ontology,wan2026tgca,wang2026hfr}. These approaches place knowledge either in the representation or directly in candidate generation and evaluation.}

\subsubsection{LLM-Guided Model Discovery}
\textcolor{red}{Recent studies use LLMs to propose and revise symbolic expressions \cite{merler2024incontext,shojaee2024llm,zhang2026llm}. LLMs also support broader search and optimization: FunSearch combines LLM generation with external evaluation to evolve programs \cite{romera2024mathematical}, while CoE-SAEA uses LLMs to evolve configuration strategies for surrogate-assisted optimization \cite{xie2025co}. These studies demonstrate that LLMs can serve as iterative search operators; their use alone is therefore not the contribution of CoSyDelay. Knowledge-guided SR constrains candidate behavior, while LLM-guided search proposes and revises expression structures. CoSyDelay integrates both within a traffic-delay-specific formulation that combines shared structure discovery, movement-specific coefficient calibration, and diagnostic feedback from estimation accuracy and traffic-delay principles.}

\subsubsection{Applications of Symbolic Regression in Traffic Systems}
\textcolor{red}{Applications of SR in traffic systems span prediction, behavior modeling, and control. For traffic prediction, Li et al.~\cite{linchao2017traffic} use genetic programming (GP) with Pareto optimization to derive highway-speed expressions, while Patelli et al.~\cite{patelli2020traffic} introduce lag operators to model temporal dependence in urban traffic flow. For behavior modeling, Li et al.~\cite{li2025discovering} use neural-network-based SR to discover explicit car-following relationships. For signal control, Liu et al.~\cite{liu2024towards} formulate phase switching as an SR problem and evolve interpretable control rules.}

\begin{table*}[!t]
\centering
\caption{\textcolor{red}{Comparison of related modeling approaches}}
\label{tab:literature_comparison}
\footnotesize
\setlength{\tabcolsep}{1.2pt}
\renewcommand{\arraystretch}{1.18}
{\color{red}
\begin{tabular}{>{\raggedright\arraybackslash}p{2.9cm}
                >{\raggedright\arraybackslash}p{2.5cm}
                >{\raggedright\arraybackslash}p{2.8cm}
                >{\raggedright\arraybackslash}p{3.0cm}
                >{\raggedright\arraybackslash}p{3.3cm}
                >{\raggedright\arraybackslash}p{3.1cm}}
\toprule
\textbf{Approach} & \textbf{Model form} & \textbf{Modeling method} & \textbf{Structure discovery} & \textbf{Model evaluation} & \textbf{Model output} \\
\midrule
Analytical delay models \cite{webster1958traffic,akcelik1981traffic,manual2010hcm2010}
& Predefined explicit equation
& Traffic-theory-based modeling
& Predefined structure
& Estimation accuracy
& Lane-group, approach, and intersection delay \\

Graph-based traffic forecasting \cite{yang2024ssgcrtn,yang2025mstdfgrn,yang2025pstcgcn,yang2025tiidgcn,yang2025sdsinet,yang2025mtegcrn,yang2025gdgcrn,yang2026dmfgcrn}
& Implicit model
& Graph-based spatiotemporal learning
& Predefined architecture with learned parameters
& Prediction accuracy
& Node- and link-level traffic states \\

Machine-learning delay models \cite{murat2006comparison,preethi2016modelling,bagdatli2021vehicle,ranpura2024estimation}
& Implicit model
& Supervised data-driven learning
& Predefined model class with learned parameters
& Estimation accuracy
& Vehicle- or intersection-level delay \\

Semi-empirical delay models \cite{korkmaz2017delay,korkmaz2020comparison}
& Predefined explicit equation
& Semi-empirical modeling
& Fixed equation with calibrated coefficients
& Estimation accuracy
& Intersection delay \\

Traffic SR \cite{linchao2017traffic,patelli2020traffic,liu2024towards,li2025discovering}
& Explicit equation
& Data-driven symbolic regression
& Operator-based search from data
& Prediction accuracy and expression complexity
& Task-specific traffic relationship \\

Knowledge-guided SR \cite{crochepierre2022reinforcement,tenachi2023deep,kronberger2022shape,bladek2023counterexample}
& Explicit equation
& Knowledge-guided symbolic regression
& Constraint-guided search from data
& Prediction accuracy and general constraints
& Task-specific equation \\

\textbf{CoSyDelay}
& \textbf{Explicit delay equation}
& \textbf{Traffic-delay-specific symbolic regression}
& \textbf{LLM-guided iterative search}
& \textbf{Estimation accuracy and traffic-delay principles}
& \textbf{Movement-, approach-, and intersection-level delay} \\
\bottomrule
\end{tabular}
}
\end{table*}

\textcolor{red}{Table~\ref{tab:literature_comparison} highlights the limitations of existing approaches. Analytical and semi-empirical models provide explicit equations but rely on predefined structures, whereas data-driven models learn flexible relationships but encode them implicitly. Traffic SR can discover explicit expressions from data, but existing applications do not jointly enforce traffic-delay principles and support multi-granularity delay estimation. Knowledge-guided SR incorporates domain constraints into symbolic search, but its existing constraints are not designed specifically for traffic delay. CoSyDelay combines explicit structure discovery, traffic-delay knowledge, and diagnostic feedback in one workflow. It searches for a shared expression structure and jointly calibrates its movement-specific coefficients against approach-delay labels, reducing the complexity of model discovery while retaining movement-specific behavior. Each candidate is evaluated for estimation accuracy and compliance with seven delay principles, and the accuracy results and diagnosed violations guide its subsequent revision. The resulting movement-level estimates are aggregated using traffic-volume weights to obtain approach- and intersection-level delays.}

% \label{Background}
\section{Method}
% \subsection{Framework}
 \begin{figure*}[!t]
        \centering
\includegraphics[width=7in]{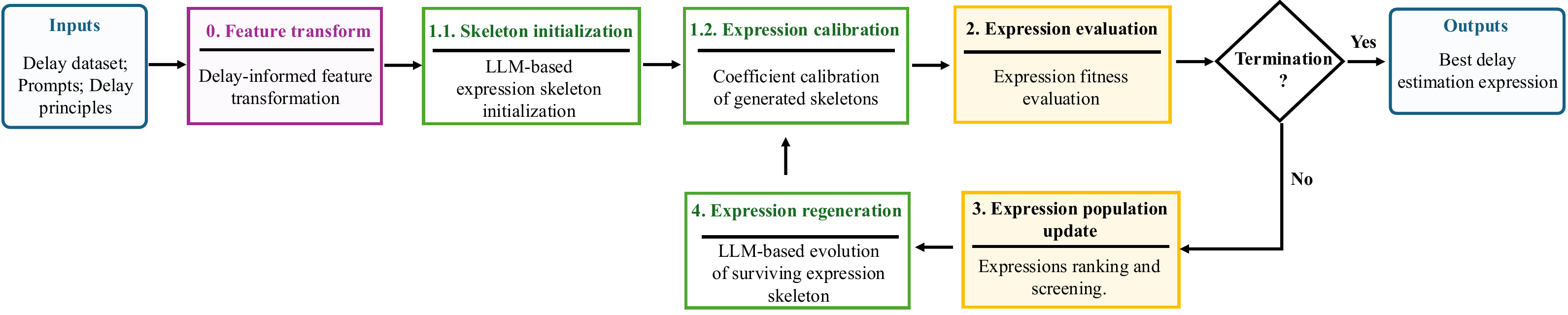}
\caption{\textcolor{red}{CoSyDelay workflow. Candidate expression structures are generated, calibrated, and evaluated; estimation accuracy and principle violations guide subsequent revisions.}}
        \label{fig:Workflow_CoSyDelay}
\end{figure*}

\textcolor{red}{Figure~\ref{fig:Workflow_CoSyDelay} illustrates the workflow, which iteratively generates, calibrates, evaluates, and revises candidate expressions. The workflow can be summarized as follows:}

\textbf{Input.} The algorithm takes as input 
\(\mathcal{I} = \big( \mathcal{DB}, \{\delta_{A,t}\}, \mathbf{p}, \mathcal{K}, \Theta \big)\),
where \(\mathcal{DB}=\{\mathcal{DB}_1,\ldots,\mathcal{DB}_n\}\) contains the movement-level inputs, with \(\mathcal{DB}_i=\{\mathbf{x}^{Turn}_{i,t}\}_{t=1}^{N_i}\) and \(\mathbf{x}^{Turn}_{i,t}=(q_{i,t},g^{\mathrm{eff}}_{i,t},C_{i,t})^\top\in\mathbb{R}^3\). Here, \(q_{i,t}\), \(g^{\mathrm{eff}}_{i,t}\), and \(C_{i,t}\) denote movement volume, effective green time, and cycle length. The dataset also provides approach-level delay labels \(\delta_{A,t}\), where \(A\) denotes an approach and \(\mathcal{M}_A\) its turning movements. These labels supervise coefficient calibration and candidate evaluation; movement-level delay labels are not required.
\(\mathbf{p}=(\mathbf{p}_{\mathrm{init}},\mathbf{p}_{\mathrm{reg}})\) denotes the LLM prompts used for initialization and regeneration, \(\mathcal{K}\) represents the domain knowledge of traffic delay principles, which is applied for physical verification, and 
\(\Theta = (s_0, G, P)\) specifies the hyperparameter set, including the per-lane saturation flow rate \(s_0\), the number of generations \(G\), and the population size \(P\).
The generation counter is initialized as \(g = 1\). We use \(s_0=1800\) veh/h/lane and define the movement saturation flow as \(s_i=s_0L_i\), where \(L_i\) is the number of lanes serving movement \(i\). Thus, the flow ratio compares the total movement volume with the corresponding movement-level saturation flow.

\begin{table*}[!t]
\centering
\caption{\textcolor{red}{Inputs and outputs of the main \textbf{CoSyDelay} components}}
\label{tab:component_dimensions}
\footnotesize
\setlength{\tabcolsep}{2pt}
\renewcommand{\arraystretch}{1.06}
{\color{red}
\begin{tabular}{>{\raggedright\arraybackslash}p{4.3cm}
                >{\raggedright\arraybackslash}p{6.1cm}
                @{\hspace{10pt}}
                >{\raggedright\arraybackslash}p{6.6cm}}
\toprule
\textbf{Component} & \textbf{Inputs and dimensions} & \textbf{Outputs and dimensions} \\
\midrule
Feature Transformation
& Raw movement observations \(\mathbf{X}_i\in\mathbb{R}^{N_i\times 3}\) for \(n\) movements, and movement saturation flows \(s_i\)
& Transformed observations \(\mathbf{X}'_i\in\mathbb{R}^{N_i\times 3}\), containing flow ratio, green ratio, and cycle length \\

Expression Initialization
& Initialization prompt, transformed variables, and \(r\) delay principles
& Initial population with \(P\) candidate expression structures \\

Expression Calibration
& A population of \(P\) candidate expression structures, transformed inputs and traffic volumes for \(n\) movements, and approach-delay labels
& \(n\times P\) calibrated movement-specific functions \\

Expression Evaluation
& \(n\times P\) calibrated functions, transformed inputs and traffic volumes, approach-delay labels, and \(r\) delay principles
& \(P\) accuracy scores, \(n\times P\) physical indicators, \(P\) overall fitness values, and \(P\times r\) principle-feedback values \\

Expression Population Update
& Parent and offspring populations (\(P\) candidate expression structures each) and their \(2P\) overall fitness values
& Updated population containing the globally best \(P\) structures and one current best expression \\

Expression Regeneration
& Updated population (\(P\) structures), \(P\) accuracy results, \(P\times r\) principle feedback, and regeneration prompt
& Offspring population with \(P\) regenerated structures \\

Multi-Granularity Delay Aggregation
& Final \(n\) movement functions, corresponding transformed inputs, and traffic volumes
& \(n\) movement estimates and their traffic-volume-weighted approach and intersection estimates \\
\midrule
\multicolumn{3}{>{\raggedright\arraybackslash}p{17.6cm}}{\footnotesize\textit{Note:} \(n\) is the number of turning movements, \(P\) is the population size, \(N_i\) is the number of samples for movement \(i\), and \(r\) is the number of traffic-delay principles.} \\
\bottomrule
\end{tabular}

}
\end{table*}

\textbf{Step 1 (Feature Transformation).} 
To facilitate SR, we define the transformed feature vector of the \(t\)-th sample for the \(i\)-th turning movement as
\(\mathbf{x}^{Turn'}_{i,t} = (y_{i,t},\lambda_{i,t},C_{i,t})^\top
= (\frac{q_{i,t}}{s_i},\frac{g^{\mathrm{eff}}_{i,t}}{C_{i,t}},C_{i,t})^\top\),
where \(y_{i,t}=q_{i,t}/s_i\) is the flow ratio and \(\lambda_{i,t}=g^{\mathrm{eff}}_{i,t}/C_{i,t}\) is the green ratio. Let \(\mathbf{X}_i=[\mathbf{x}^{Turn}_{i,1},\ldots,\mathbf{x}^{Turn}_{i,N_i}]^\top\in\mathbb{R}^{N_i\times3}\) and \(\mathbf{X}'_i=[\mathbf{x}^{Turn'}_{i,1},\ldots,\mathbf{x}^{Turn'}_{i,N_i}]^\top\in\mathbb{R}^{N_i\times3}\). The resulting transformed dataset is \(\mathcal{DB}'_i=\{\mathbf{x}^{Turn'}_{i,t}\}_{t=1}^{N_i}\). The flow and green ratios express demand relative to saturation flow and green allocation relative to cycle length. These traffic-theory variables help the search construct meaningful delay expressions without changing the number of input variables.

\begin{figure*}[!t]
        \centering
\includegraphics[height=3.65 in,width=7in]{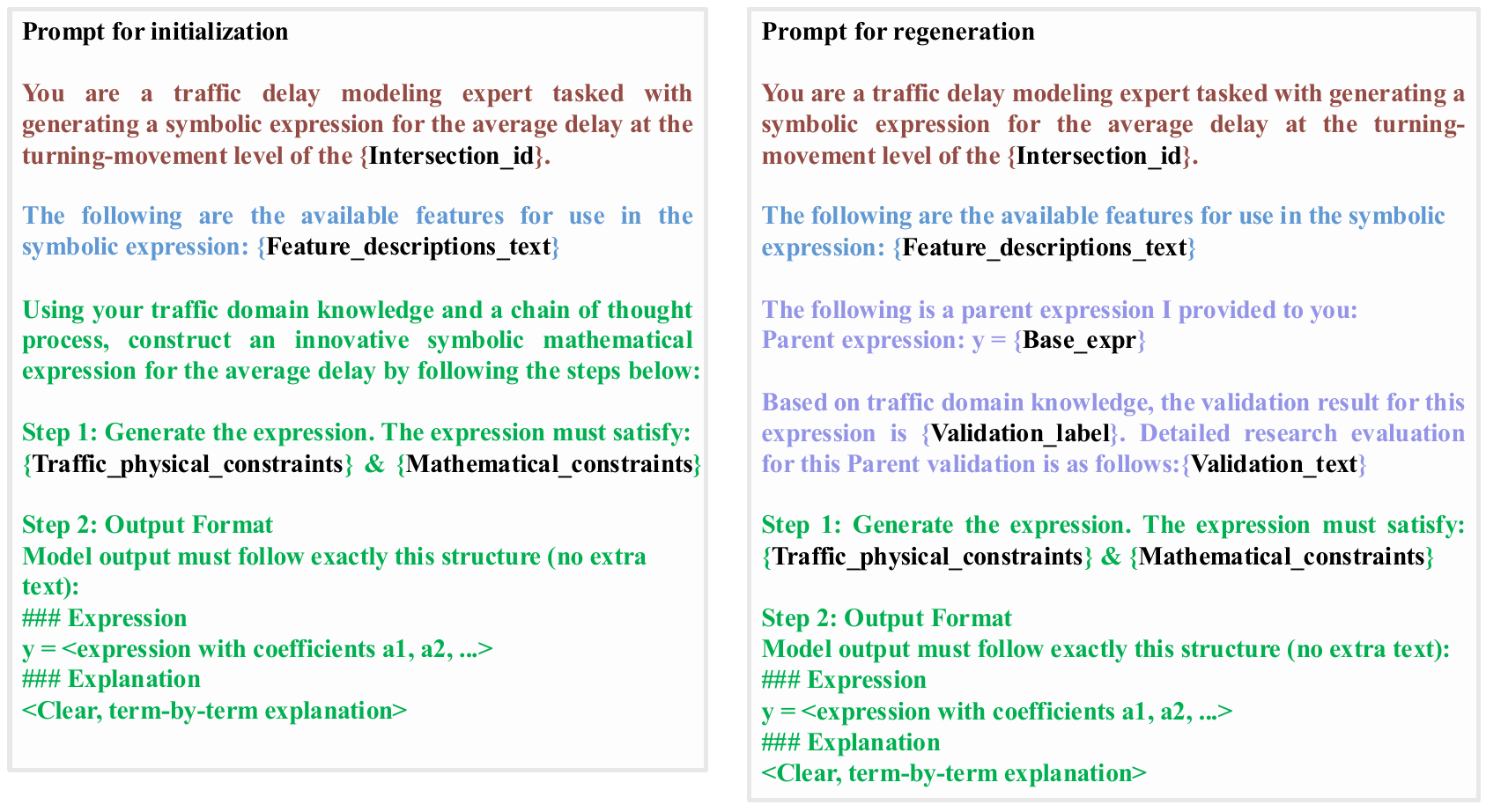}
        \caption{Initialization (left) and regeneration (right) prompts in CoSyDelay. Colors distinguish the role, input context, task instructions, and evaluation feedback.}
        \label{fig:Prompt_CoSyDelay}
\end{figure*}

\textbf{Step 2 (Initialization of Symbolic Expressions).} 
At generation \(g=1\), the LLM is prompted, using the initialization prompt in Fig.~\ref{fig:Prompt_CoSyDelay}, to generate an initial population of \(P\) symbolic delay expressions, denoted by \(\mathcal{P}^{(1)} = \{\phi_j^{(1)}\}_{j=1}^{P}\). Each expression consists of traffic-related variables and symbolic coefficients \(a_1, a_2, \dots, a_k\), without assigning fixed numerical values to these coefficients. Each movement has its own coefficient vector \(\mathbf{a}_{i,j}\). For each candidate structure, these vectors are calibrated against the training approach-delay labels by minimizing the discrepancy between approach-level delay labels and the volume-weighted aggregation of movement estimates. Thus, the parameters are movement-specific, but calibration is supervised at the approach level. The resulting movement-specific function is \(f_{i,j}^{(1)}=\phi_j^{(1)}(\cdot;\mathbf{a}_{i,j})\). The instantiated candidate-function set induced by \(\mathcal{P}^{(1)}\) is thus \(\mathcal{F}^{(1)} = \{ f_{i,j}^{(1)} \mid i = 1, \dots, n;\, j = 1, \dots, P \}\).

\textbf{Step 3 (Evaluation).} 
For each symbolic expression \(\phi_j^{(g)}\), the calibrated movement estimates are first aggregated into approach-delay estimates:
\[
f_{A,j}^{(g)}(\mathbf{x}^{Approach\prime}_{A,t})=\frac{\sum_{i\in\mathcal{M}_A}q_{i,t}f_{i,j}^{(g)}(\mathbf{x}^{Turn'}_{i,t})}{\sum_{i\in\mathcal{M}_A}q_{i,t}}.
\]
Only samples with positive approach volume are included. Candidate Fitness combines approach-level estimation accuracy and physical consistency:
\begin{equation}
\mathrm{Fitness}_{\mathrm{overall}}(\phi_j^{(g)})=\mathrm{Score}_{\mathrm{acc}}(\phi_j^{(g)})+\mathrm{Score}_{\mathrm{phys}}(\phi_j^{(g)}).
\label{eq:fitness_function}
\end{equation}
Here, \(f_{A,j}^{(g)}\) is the approach-delay function obtained by aggregating movement-level delay estimates using traffic-volume weights. Its input \(\mathbf{x}^{Approach\prime}_{A,t}\) includes the transformed features and corresponding traffic volumes of all turning movements in approach \(A\). Let \(\mathcal{A}_{\mathrm{valid}}\) denote the set of approaches with positive total volume and nonzero delay-label variance, and let \(m=|\mathcal{A}_{\mathrm{valid}}|\). The accuracy score is
\begin{equation}
\begin{aligned}
&\mathrm{Score}_{\mathrm{acc}}(\phi_j^{(g)})=\max\Bigg\{0,\\[2pt]
&\quad\frac{1}{m}\sum_{A\in\mathcal{A}_{\mathrm{valid}}}
\left[1-\frac{\sum_t\left(\delta_{A,t}-f_{A,j}^{(g)}(\mathbf{x}^{Approach\prime}_{A,t})\right)^2}
{\sum_t(\delta_{A,t}-\bar\delta_A)^2}\right]\Bigg\}.
\end{aligned}
\end{equation}
Here, \(\bar\delta_A\) is the mean delay label at approach \(A\), and the sums use its valid evaluation samples. The approach-level coefficients of determination (\(R^2\)) are first averaged; the accuracy score equals this mean if it is nonnegative and is set to zero otherwise. During search, coefficient calibration and population updates use training data; validation Fitness selects the final expression.

{\color{red}
The movement-level physical indicator \(s_{i,j}^{\mathrm{phys},(g)}\) evaluates the candidate function \(f_{i,j}^{(g)}\) against the delay principles in Table~\ref{tab:traffic_physics_rules}. Let \(y=q_i/s_i\), where \(s_i\) is the corresponding movement saturation flow, denote the movement flow ratio and \(\lambda=g^{\mathrm{eff}}/C\) the corresponding green ratio. Principle~1 checks for the required variables. Principles~2 and~3 examine monotonicity by differentiating the expression with respect to \(y\) and \(\lambda\), respectively. Principle~4 tests unit consistency through first-order homogeneity in \(C\). Principles~5 and~7 use symbolic one-sided limits to examine low-flow and zero-green behavior, while Principle~6 evaluates the validity of the resulting delay values. The numerical checks in Principles~2, 3, and~6 are conducted over \(\mathcal{D}=[0.001,2.0]\times[0.02,0.95]\times[30,240]\) for \((y,\lambda,C)\), with \(C\) measured in seconds. This domain contains the observed training ranges and extends beyond them to provide conservative stress-test conditions. It is sampled using a fixed space-filling Latin hypercube design \cite{mckay1979comparison}, supplemented by combinations of the variable bounds. Let \(v_{i,j,l}^{(g)}=1\) if function \(f_{i,j}^{(g)}\) passes principle \(l\), and zero otherwise. Formally,}
\begin{equation}
s_{i,j}^{\mathrm{phys},(g)}=
\begin{cases}
1, & \text{if all principles pass},\\
0, & \text{otherwise}.
\end{cases}
\end{equation}
A movement-level indicator of one means that the function passes all specified principle checks.
\textcolor{red}{For feedback-guided regeneration, the principle-level verification results are summarized by \(\mathbf{V}^{(g)}\in\{0,1\}^{P\times r}\), where \(V_{j,l}^{(g)}=\prod_{i=1}^{n}v_{i,j,l}^{(g)}\). Thus, \(V_{j,l}^{(g)}=1\) only when every movement-specific instance of \(\phi_j^{(g)}\) satisfies principle \(l\), and the LLM receives the specific principles violated by each symbolic structure rather than only its aggregate physical-consistency score.}
The candidate-level physical score requires every calibrated movement function to pass:
\begin{equation}
\mathrm{Score}_{\mathrm{phys}}(\phi_j^{(g)})=
\begin{cases}
1, & \text{if all movements pass},\\
0, & \text{otherwise}.
\end{cases}
\end{equation}
The resulting overall Fitness lies in \([0,2]\) and is used to rank candidate expression structures.

\begin{table*}[htbp]
\centering
\caption{\textcolor{red}{Traffic-delay principles and automated checks}}
\label{tab:traffic_physics_rules}
\footnotesize
\setlength{\tabcolsep}{3pt}
\renewcommand{\arraystretch}{1.12}
{\color{red}
\begin{tabular}{>{\raggedright\arraybackslash}p{3.1cm}
                >{\raggedright\arraybackslash}p{7.0cm}
                >{\raggedright\arraybackslash}p{7.0cm}}
\toprule
\textbf{Principle} & \textbf{Requirement} & \textbf{Automated check} \\
\midrule
1. Required Traffic Variables
& \(f\) must include the movement flow ratio \(y\), green ratio \(\lambda\), and cycle time \(C\).
& Confirm that \(y\), \(\lambda\), and \(C\) are present in the expression. \\

2. Monotonicity w.r.t. Flow Ratio
& Delay must be non-decreasing in \(y\): \(\partial f/\partial y\geq0\).
& Compute \(\partial f/\partial y\) and confirm that every sampled value is real, finite, and non-negative. \\

3. Monotonicity w.r.t. Green Ratio
& Delay must be non-increasing in \(\lambda\): \(\partial f/\partial\lambda\leq0\).
& Compute \(\partial f/\partial\lambda\) and confirm that every sampled value is real, finite, and non-positive. \\

4. Unit Consistency
& With dimensionless \(y\), \(\lambda\), and coefficients, \(f(y,\lambda,kC)=k f(y,\lambda,C)\) for \(k>0\).
& Verify symbolically that \(f(y,\lambda,kC)-k f(y,\lambda,C)\equiv0\). \\

5. Low-Flow Condition
& For \(\lambda,C>0\), \(\lim_{y\to0^+}f(y,\lambda,C)\) must be finite and non-negative, but need not be zero.
& Compute the symbolic limit and confirm that it exists and lies in \([0,\infty)\). \\

6. Non-Negativity
& Delay must remain real, finite, and non-negative over the verification domain.
& Evaluate \(f\) and confirm that every sampled value is real, finite, and non-negative. \\

7. Zero Green Limit
& For \(y,C>0\), \(\lim_{\lambda\to0^+}f(y,\lambda,C)=+\infty\).
& Compute the symbolic limit and confirm that it equals \(+\infty\) for \(y,C>0\). \\
\bottomrule
\end{tabular}
}
\end{table*}

\textbf{Step 4 (Population Update and Termination Check).}
For \(g=1\), the initialized population is directly taken as the current population after evaluation, namely \(\mathcal{P}^{(1)}\). For \(g>1\), the offspring population \(\mathcal{P}^{\mathrm{off},(g)}\) is first evaluated in the same manner as in Step~3. The parent and offspring populations are pooled, and the \(P\) structures with the highest \(\mathrm{Fitness}_{\mathrm{overall}}\) values are retained as \(\mathcal{P}^{(g)}\). This is global parent--offspring survivor selection (the standard \((\mu+\lambda)\) scheme), using \(P\) parents and \(P\) offspring. The best symbolic expression in the current population is subsequently identified as \(\tilde{\phi}^{(g)} = \arg\max_{\phi \in \mathcal{P}^{(g)}} \mathrm{Fitness}_{\mathrm{overall}}(\phi)\). If \(g=G\), the evolutionary process terminates. Otherwise, the algorithm proceeds to Step~5 to generate the offspring population for generation \(g+1\).

% The generalized expressions in \(\mathcal{P}^{(g)}\) are first ranked according to their overall fitness values. The best expression is selected as the current parent,
% $\tilde{\phi}^{(g)} = \arg\max_{\phi \in \mathcal{P}^{(g)}} \mathrm{Fitness}_{\mathrm{overall}}(\phi),$
% which provides the reference symbolic structure for offspring generation in the next step.

\textbf{Step 5 (Regeneration).} 
If the termination condition is not met, each parent symbolic expression \(\phi_j^{(g)} \in \mathcal{P}^{(g)}\) is used to generate one corresponding offspring expression for generation \(g+1\). Specifically, \textbf{CoSyDelay} inserts \(\phi_j^{(g)}\) together with its training-set estimation-accuracy results and principle-level verification feedback from Table~\ref{tab:traffic_physics_rules} into the regeneration prompt shown in Fig.~\ref{fig:Prompt_CoSyDelay}, and the LLM returns a regenerated symbolic variant \(\phi_j^{\mathrm{off},(g+1)}\). Repeating this process for all \(j=1,\dots,P\) yields \(\mathcal{P}^{\mathrm{off},(g+1)}=\{\phi_j^{\mathrm{off},(g+1)}\}_{j=1}^{P}\). The movement-specific coefficient vectors are then jointly calibrated against the training approach-delay labels. Finally, the counter is updated as \(g\gets g+1\), and the instantiated offspring functions are passed to Step~3 for evaluation.

\textbf{Output.}  
After evolution terminates, the shared expression structure with the highest validation Fitness is selected from the final population. Let
\(j^*=\arg\max_{j\in\{1,\ldots,P\}}\mathrm{Fitness}_{\mathrm{val}}(\phi_j^{(G)})\)
denote its index, where validation Fitness uses the same definition as Eq.~\ref{eq:fitness_function} and is evaluated using the training-fitted coefficients without refitting on validation data. Only after selection, with the structure fixed, are movement-specific coefficients refitted on the combined training and validation sets. The algorithm outputs the corresponding movement-specific delay functions \(\{f_i^*\}_{i=1}^n\), where \(f_i^*(\cdot)=\phi_{j^*}^{(G)}(\cdot;\mathbf{a}_i^*)\) and \(\mathbf{a}_i^*\) denotes the refitted coefficient vector for movement \(i\). Thus, the output functions share one expression structure but contain movement-specific coefficients jointly calibrated against approach-delay labels. \textcolor{red}{Only these movement-level functions are learned by SR; approach- and intersection-level delays are subsequently computed from their estimates.}

{\color{red}
Following the standard aggregation of lane-group and movement control delays in the Highway Capacity Manual and Federal Highway Administration (FHWA) traffic signal timing guidance \cite{manual2010hcm2010,fhwa2008traffic}, we compute approach- and intersection-level mean delays using traffic-volume-weighted averages. Let \(\hat d_i=f_i^*(\mathbf{x}^{Turn'}_i)\) denote the estimated mean delay per vehicle for movement \(i\) during a common analysis period, and let \(q_i\) denote its traffic volume. For an approach \(A\) with constituent movement set \(\mathcal{M}_A\) and total volume \(Q_A=\sum_{i\in\mathcal{M}_A}q_i>0\), the approach-level mean delay is
\begin{align}
\hat d_A
&=\frac{\sum_{i\in\mathcal{M}_A}q_i\hat d_i}{Q_A}
=\frac{\sum_{i\in\mathcal{M}_A}q_i f_i^*(\mathbf{x}^{Turn'}_i)}{\sum_{i\in\mathcal{M}_A}q_i}.
\end{align}
Similarly, for an intersection with approach set \(\mathcal{A}\) and positive total volume, the intersection-level mean delay is
\begin{align}
\hat d_{\mathrm{int}}
&=\frac{\sum_{A\in\mathcal{A}}Q_A\hat d_A}{\sum_{A\in\mathcal{A}}Q_A}
=\frac{\sum_{A\in\mathcal{A}}\sum_{i\in\mathcal{M}_A}q_i\hat d_i}
{\sum_{A\in\mathcal{A}}\sum_{i\in\mathcal{M}_A}q_i}.
\end{align}
These equations do not constitute independently learned higher-level models; they follow directly from the definition of mean per-vehicle delay. Their use assumes that all movement estimates employ the same delay definition and analysis period, that the movement sets are mutually exclusive and collectively cover the vehicles being aggregated, and that the traffic-volume weights refer to that same period. Aggregation is performed only when the corresponding total volume is positive. When the total traffic volume is zero, no vehicles are available for aggregation; the mean delay therefore cannot be computed and is reported as not applicable (N/A).

Like classical delay models such as Webster, Ak\c{c}elik, and HCM \cite{webster1958traffic,akcelik1981traffic,manual2010hcm2010}, CoSyDelay uses traffic demand and signal timing as the basis for delay estimation. It learns an explicit expression from data rather than imposing a predefined delay formula, with movement-specific coefficients jointly calibrated against approach-delay labels. However, traffic-volume weighting only aggregates movement estimates; it does not model their interactions. Short turning lanes and queue spillback can obstruct adjacent movements \cite{fhwa2008traffic}, and the current inputs do not explicitly represent lane storage, shared-lane blocking, or queue interactions. Incorporating geometric and queue-state variables is therefore a direction for future work.}

\begin{figure}[!htbp]
        \centering
\includegraphics[height=1.55 in,width=3.35in]{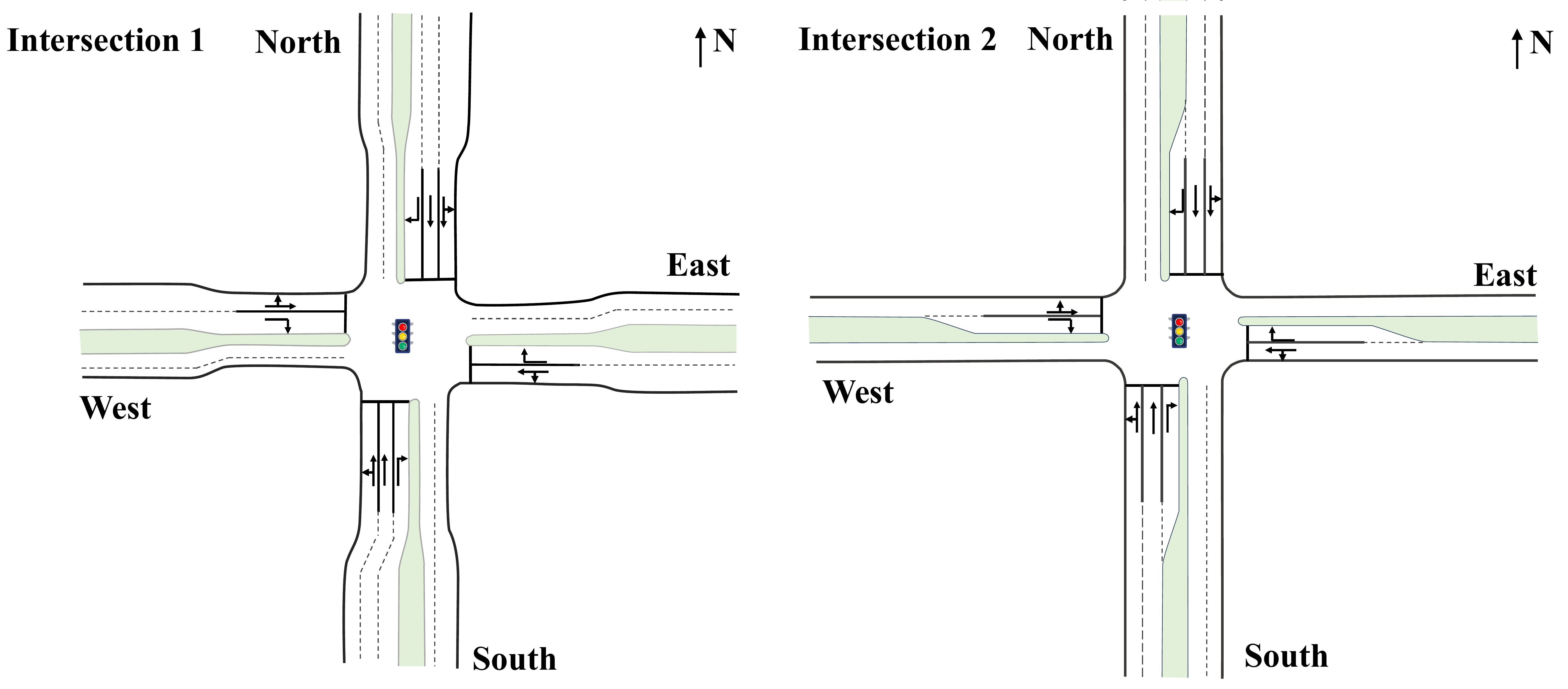}
\caption{Geometrical layout of Intersection 1 (left) and Intersection 2 (right).}
        \label{fig:Inter_1_2}
\end{figure}
\begin{figure*}[!t]
        \centering
\includegraphics[height=1.7 in,width=7in]{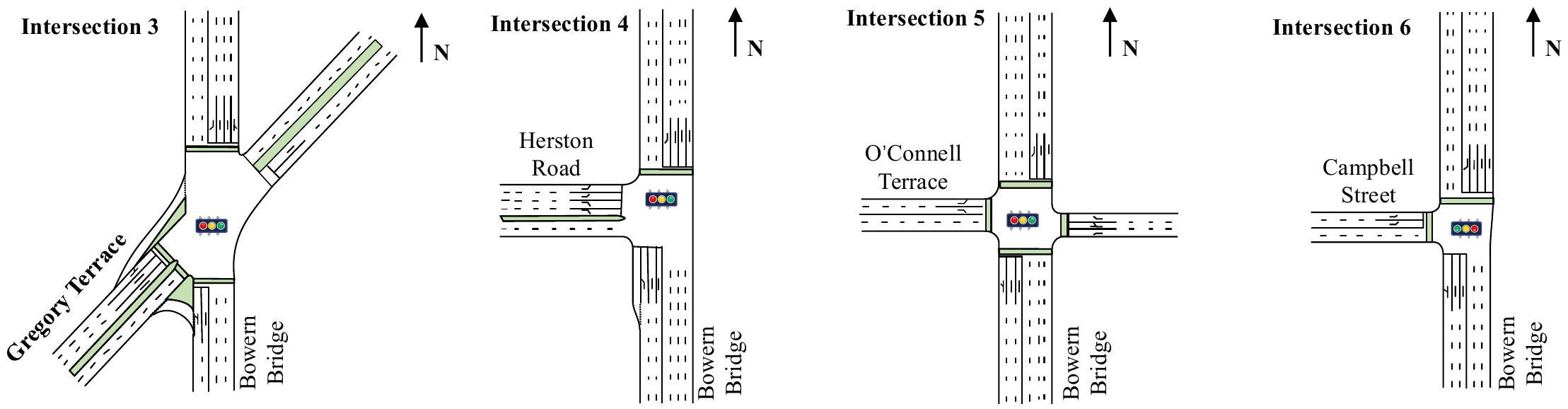}
\caption{Geometrical layout of Intersections 3 to 6, from left to right corresponding to Intersections 3, 4, 5, and 6.}
        \label{fig:Inter_3_6}
\end{figure*}
\begin{table*}[t]
\centering
\caption{Intersection configurations and phase control. S, E, N, and W denote the south, east, north, and west approaches; L, T, and R denote left, through, and right movements. Each phase lists its served approach--movement pairs.}
\renewcommand{\arraystretch}{1.05}
\setlength{\tabcolsep}{4pt}
\footnotesize
\begin{tabular}{c c p{4.2cm} p{8.0cm}}

\toprule
\textbf{Intersection} & \textbf{Intergreen (s)} & \textbf{Movements by approach} & \textbf{Phase allocation} \\ 
\midrule
1--2 & 24 & S/E/N/W: L, T, R &
\makecell[l]{A: S/N $\to$ R; B: S/N $\to$ L, T; \\
C: E/W $\to$ R; D: E/W $\to$ L, T} \\
\midrule
3 & 18 & 
S: L, T; E: L, T; N: L, T, R; W: L, T &
\makecell[l]{A: S/N $\to$ L, T; W $\to$ L; B: N $\to$ L, T, R; W $\to$ L; \\
C: E/W $\to$ L, T} \\
\midrule
4 & 18 & 
S: L, T; N: T, R; W: L, R &
\makecell[l]{A: S $\to$ L, T; N $\to$ T; B: S $\to$ L; N $\to$ T, R; W $\to$ L; \\
C: S $\to$ L; W $\to$ L, R} \\
\midrule
5 & 18 & 
S: L, T; E: L, T, R; N: T; W: L, R &
\makecell[l]{A: E/W $\to$ L, R; B: E $\to$ L, T, R; W $\to$ L; \\
C: S $\to$ L, T; N $\to$ T; W $\to$ L} \\
\midrule
6 & 12 & 
S: T, R; E: L; N: L, T &
\makecell[l]{A: S $\to$ T; N $\to$ L, T; \\
B: S $\to$ T, R; E $\to$ L} \\
\bottomrule

\end{tabular}
\label{tab:phases_for_intersections}
\end{table*}

\section{Numerical Experiments}
\subsection{Experimental Setup}
Experiments use six signalized intersections. Intersections~1 and~2 are \textcolor{red}{simulated} standard four-leg intersections (Fig.~\ref{fig:Inter_1_2}), with short right-turn lanes at Intersection~2. \textcolor{red}{For Intersections~3--6, located along Bowen Bridge Road in Queensland, Australia (Fig.~\ref{fig:Inter_3_6}), field-collected traffic volumes and signal timings are used in simulation to obtain the delay labels.} \textcolor{revisionRtwo}{Delay labels measure time loss within the approach detection area relative to a reference speed determined by road speed limits and vehicle speed settings. Simulations start from an empty network and last 4,500~s: 900~s of warm-up and 3,600~s of measurement (900--4,500~s). Only in-window time loss is counted, including contributions from vehicles entering during warm-up or remaining at 4,500~s. Vehicles and queues are retained after warm-up. Simulations are not extended to clear queues; post-window delay and waiting before network insertion are excluded.} \textcolor{revisionRtwo}{The available delay labels are approach-level averages. Movement-specific coefficients are therefore jointly calibrated using approach-level delay labels, and estimation accuracy is evaluated at the approach level.} Table~\ref{tab:phases_for_intersections} summarizes their movement configurations and signal phases. \textcolor{revisionRtwo}{The dataset contains 10,274 samples: 5,909 training samples for coefficient fitting and evolutionary search, 1,512 validation samples for final expression selection, and 2,853 test samples for final evaluation.} \textcolor{red}{ All methods use the same fixed splits. Hyperparameters of the data-driven baselines are selected using only the validation split.}
Estimation accuracy is measured using \(R^2\), root mean squared error (RMSE), and mean absolute error (MAE). When physical consistency is also considered, Fitness is calculated using Eq.~\ref{eq:fitness_function}. \textcolor{red}{Each stochastic method is evaluated over 20 independent runs using the same data splits and method-specific settings, and its results are reported as mean \(\pm\) standard deviation. Deterministic analytical methods are reported once for each setting. Average Rank summarizes performance over the evaluation cases in each table or experimental block; lower values are better. Statistical significance is evaluated using two-sided one-sample Wilcoxon signed-rank tests on the paired performance differences, with Holm-adjusted \(p\)-values at \(\alpha=0.05\). The symbols \(+\), \(=\), and \(-\) indicate that CoSyDelay is significantly better, not significantly different, or significantly worse than the comparison method, respectively.}
\textbf{CoSyDelay} uses a population size of \(P=20\), \(G=10\) generations, and a per-lane saturation flow of \(s_0=1800\) veh/h/lane. The default LLM is GPT-4.1-mini with temperature \(0.0\), top-\(p=1.0\), a 1024-token output limit, and a 30-s timeout. \textcolor{revisionRtwo}{At initialization and regeneration, the LLM returns one candidate per call as Expression and Explanation text (Fig.~\ref{fig:Prompt_CoSyDelay}). Candidates are rejected if the expression cannot be extracted, parsed, or evaluated, contains unsupported syntax, variables, or operators, or duplicates an existing candidate.} \textcolor{red}{ Failed calls are retried, while the rejection reason for an invalid or duplicate candidate is returned to the LLM for revision. If no valid, unique expression is obtained after five attempts, candidate generation is restarted with a new initialization.} Experiments were conducted on an Intel Core i7-10700K central processing unit (CPU) and an NVIDIA RTX 4080 Super graphics processing unit (GPU).

\subsection{Estimation Accuracy Comparison with Baseline Methods}

\textcolor{red}{We compare CoSyDelay with two groups of baselines. The first group contains the standard and data-calibrated forms of three analytical delay models: Webster~\cite{webster1958traffic}, HCM~\cite{manual2010hcm2010}, and Ak\c{c}elik~\cite{akcelik1981traffic}. The second group contains data-driven methods: linear regression, random forest (RF), Extreme Gradient Boosting (XGBoost)~\cite{chen2016xgboost}, AdaBoost, support vector regression (SVR), k-nearest neighbors (KNN), graph convolutional network (GCN), GRANDE~\cite{marton2024grande}, ModernNCA~\cite{ye2025modernnca}, Dirichlet graph auto-encoder (DGAE)~\cite{zhou2025dgae}, and Graph Neural Network for Urban Interpolation (GNNUI)~\cite{kaiser2025gnnui}. All methods are evaluated using $R^2$, RMSE, and MAE.}

\begin{table*}[t]
\centering
\color{red}
\scriptsize
\setlength{\tabcolsep}{2.5pt}
\caption{Test performance of analytical delay models and CoSyDelay at I1--I6. Metrics are averaged across valid approaches. Standard analytical formulas are deterministic; calibrated variants and CoSyDelay are reported as mean $\pm$ standard deviation over 20 runs. Average Rank averages ranks across intersections and metrics. Superscripts summarize Holm-adjusted two-sided one-sample Wilcoxon signed-rank tests on the paired performance differences ($\alpha=0.05$): $+$, $=$, and $-$ denote CoSyDelay performing significantly better, not significantly differently, and significantly worse, respectively. The final row counts these outcomes; best values are bold.}
\label{tab:analytical_cosydelay}
\begin{tabular*}{0.90\textwidth}{@{\extracolsep{\fill}}llccccccc@{}}
\toprule
Site & Metric & Webster & Webster-opt & HCM & HCM-opt & Ak\c{c}elik & Ak\c{c}elik-opt & CoSyDelay \\
\midrule
    \multirow{3}{*}{I1} & $R^2$ & -0.367$^{+}$ & \meanstd{-0.834}{1.730}$^{+}$ & -0.368$^{+}$ & \meanstd{0.529}{0.277}$^{+}$ & -0.368$^{+}$ & \meanstd{-0.834}{1.730}$^{+}$ & \meanstdbf{0.803}{0.002} \\
    & RMSE & 12.294$^{+}$ & \meanstd{11.381}{4.968}$^{+}$ & 12.297$^{+}$ & \meanstd{6.435}{1.524}$^{+}$ & 12.297$^{+}$ & \meanstd{11.381}{4.968}$^{+}$ & \meanstdbf{4.527}{0.027} \\
    & MAE & 10.522$^{+}$ & \meanstd{9.021}{4.584}$^{+}$ & 10.527$^{+}$ & \meanstd{4.701}{1.242}$^{+}$ & 10.527$^{+}$ & \meanstd{9.021}{4.584}$^{+}$ & \meanstdbf{3.288}{0.019} \\
    \multirow{3}{*}{I2} & $R^2$ & -0.516$^{+}$ & \meanstd{-0.051}{0.544}$^{+}$ & -0.516$^{+}$ & \meanstd{0.510}{0.175}$^{+}$ & -0.516$^{+}$ & \meanstd{-0.051}{0.544}$^{+}$ & \meanstdbf{0.756}{0.003} \\
    & RMSE & 15.220$^{+}$ & \meanstd{11.627}{2.170}$^{+}$ & 15.220$^{+}$ & \meanstd{8.223}{1.251}$^{+}$ & 15.220$^{+}$ & \meanstd{11.627}{2.170}$^{+}$ & \meanstdbf{5.794}{0.045} \\
    & MAE & 12.526$^{+}$ & \meanstd{8.987}{2.099}$^{+}$ & 12.526$^{+}$ & \meanstd{5.905}{1.102}$^{+}$ & 12.526$^{+}$ & \meanstd{8.987}{2.099}$^{+}$ & \meanstdbf{4.160}{0.053} \\
    \multirow{3}{*}{I3} & $R^2$ & 0.191$^{+}$ & \meanstd{-0.664}{1.098}$^{+}$ & 0.222$^{+}$ & \meanstd{0.322}{0.558}$^{+}$ & 0.222$^{+}$ & \meanstd{-0.664}{1.098}$^{+}$ & \meanstdbf{0.866}{0.024} \\
    & RMSE & 10.815$^{+}$ & \meanstd{12.893}{4.413}$^{+}$ & 10.539$^{+}$ & \meanstd{8.410}{2.606}$^{+}$ & 10.539$^{+}$ & \meanstd{12.893}{4.413}$^{+}$ & \meanstdbf{4.416}{0.374} \\
    & MAE & 8.227$^{+}$ & \meanstd{10.443}{4.299}$^{+}$ & 8.131$^{+}$ & \meanstd{6.531}{2.337}$^{+}$ & 8.131$^{+}$ & \meanstd{10.443}{4.299}$^{+}$ & \meanstdbf{3.102}{0.279} \\
    \multirow{3}{*}{I4} & $R^2$ & -7.654$^{+}$ & \meanstd{-9.147}{4.840}$^{+}$ & 0.448$^{+}$ & \meanstd{-2.245}{1.674}$^{+}$ & 0.448$^{+}$ & \meanstd{-9.147}{4.840}$^{+}$ & \meanstdbf{0.861}{0.023} \\
    & RMSE & 6.926$^{+}$ & \meanstd{10.197}{3.518}$^{+}$ & 3.742$^{+}$ & \meanstd{5.921}{2.905}$^{+}$ & 3.742$^{+}$ & \meanstd{10.197}{3.518}$^{+}$ & \meanstdbf{2.070}{0.172} \\
    & MAE & 3.725$^{+}$ & \meanstd{8.873}{3.614}$^{+}$ & 2.862$^{+}$ & \meanstd{4.606}{2.969}$^{+}$ & 2.862$^{+}$ & \meanstd{8.873}{3.614}$^{+}$ & \meanstdbf{1.289}{0.112} \\
    \multirow{3}{*}{I5} & $R^2$ & -1.003$^{+}$ & \meanstd{-1.512}{1.348}$^{+}$ & -0.167$^{+}$ & \meanstd{0.598}{0.043}$^{+}$ & -0.167$^{+}$ & \meanstd{-1.512}{1.348}$^{+}$ & \meanstdbf{0.872}{0.008} \\
    & RMSE & 15.309$^{+}$ & \meanstd{14.589}{3.772}$^{+}$ & 12.001$^{+}$ & \meanstd{7.185}{0.360}$^{+}$ & 12.001$^{+}$ & \meanstd{14.589}{3.772}$^{+}$ & \meanstdbf{3.902}{0.143} \\
    & MAE & 9.299$^{+}$ & \meanstd{11.602}{4.508}$^{+}$ & 9.107$^{+}$ & \meanstd{5.428}{0.337}$^{+}$ & 9.107$^{+}$ & \meanstd{11.602}{4.508}$^{+}$ & \meanstdbf{2.651}{0.144} \\
    \multirow{3}{*}{I6} & $R^2$ & -3.048$^{+}$ & \meanstd{-4.234}{1.728}$^{+}$ & 0.087$^{+}$ & \meanstd{0.163}{0.373}$^{+}$ & 0.087$^{+}$ & \meanstd{-4.234}{1.728}$^{+}$ & \meanstdbf{0.839}{0.037} \\
    & RMSE & 7.868$^{+}$ & \meanstd{8.315}{1.525}$^{+}$ & 4.717$^{+}$ & \meanstd{4.089}{1.083}$^{+}$ & 4.717$^{+}$ & \meanstd{8.315}{1.525}$^{+}$ & \meanstdbf{2.605}{0.221} \\
    & MAE & 4.668$^{+}$ & \meanstd{7.052}{1.580}$^{+}$ & 3.906$^{+}$ & \meanstd{2.954}{0.890}$^{+}$ & 3.906$^{+}$ & \meanstd{7.052}{1.580}$^{+}$ & \meanstdbf{1.660}{0.180} \\
\midrule
\multicolumn{2}{l}{Average Rank} & 5.111 & 5.611 & 4.139 & 2.389 & 4.139 & 5.611 & \textbf{1.000} \\
\multicolumn{2}{l}{$+/=/-$} & 18/0/0 & 18/0/0 & 18/0/0 & 18/0/0 & 18/0/0 & 18/0/0 & -- \\
\bottomrule
\end{tabular*}
\end{table*}

\providecommand{\replytablebegin}{\begin{table*}[t]\centering}
\providecommand{\replytableend}{\end{table*}}
\providecommand{\replytablecaption}[1]{\caption{#1}}
\replytablebegin
  \color{red}
  \footnotesize
  \setlength{\tabcolsep}{2.5pt}
  \replytablecaption{Test performance of data-driven models and CoSyDelay at I1--I6. Metrics are averaged across valid approaches. Values are mean $\pm$ standard deviation over 20 runs; deterministic models are reported once. Average Rank averages ranks across intersections and metrics. Superscripts summarize Holm-adjusted two-sided one-sample Wilcoxon signed-rank tests on the paired performance differences ($\alpha=0.05$): $+$, $=$, and $-$ denote CoSyDelay performing significantly better, not significantly differently, and significantly worse, respectively. The final row counts these outcomes; best values are bold.}
  \ifdefined\SuppressTableVILabel\else
  \label{tab:datadriven-subset12-manuscript}
  \fi
  \resizebox{\textwidth}{!}{%
  \begin{tabular}{@{}llcccccccccccc@{}}
    \toprule
    Site & Metric & Linear & RF & XGBoost & AdaBoost & SVR & KNN & GCN & DGAE & GNNUI & GRANDE & ModernNCA & CoSyDelay \\
    \midrule
    \multirow{3}{*}{I1} & $R^2$ & 0.784$^{+}$ & \meanstd{0.777}{0.001}$^{+}$ & \meanstd{0.759}{0.002}$^{+}$ & \meanstd{0.716}{0.003}$^{+}$ & 0.760$^{+}$ & 0.715$^{+}$ & \meanstd{0.697}{0.004}$^{+}$ & \meanstd{0.747}{0.007}$^{+}$ & \meanstd{0.778}{0.006}$^{+}$ & \meanstd{0.763}{0.006}$^{+}$ & \meanstd{0.785}{0.005}$^{+}$ & \meanstdbf{0.803}{0.002} \\
    & RMSE & 4.671$^{+}$ & \meanstd{4.825}{0.014}$^{+}$ & \meanstd{4.968}{0.034}$^{+}$ & \meanstd{5.485}{0.034}$^{+}$ & 5.021$^{+}$ & 5.522$^{+}$ & \meanstd{5.666}{0.035}$^{+}$ & \meanstd{5.188}{0.073}$^{+}$ & \meanstd{4.802}{0.072}$^{+}$ & \meanstd{5.000}{0.076}$^{+}$ & \meanstd{4.703}{0.066}$^{+}$ & \meanstdbf{4.527}{0.027} \\
    & MAE & 3.363$^{+}$ & \meanstd{3.401}{0.011}$^{+}$ & \meanstd{3.529}{0.028}$^{+}$ & \meanstd{4.161}{0.032}$^{+}$ & 3.387$^{+}$ & 3.872$^{+}$ & \meanstd{3.989}{0.048}$^{+}$ & \meanstd{3.673}{0.053}$^{+}$ & \meanstd{3.395}{0.047}$^{+}$ & \meanstd{3.553}{0.057}$^{+}$ & \meanstd{3.316}{0.039}$^{+}$ & \meanstdbf{3.288}{0.019} \\
    \multirow{3}{*}{I2} & $R^2$ & 0.732$^{+}$ & \meanstd{0.745}{0.002}$^{+}$ & \meanstd{0.735}{0.004}$^{+}$ & \meanstd{0.638}{0.007}$^{+}$ & 0.728$^{+}$ & 0.694$^{+}$ & \meanstd{0.646}{0.012}$^{+}$ & \meanstd{0.747}{0.007}$^{+}$ & \meanstd{0.763}{0.011}$^{-}$ & \meanstd{0.752}{0.008}$^{=}$ & \meanstdbf{0.764}{0.017}$^{-}$ & \meanstd{0.756}{0.003} \\
    & RMSE & 6.176$^{+}$ & \meanstd{6.002}{0.023}$^{+}$ & \meanstd{6.056}{0.050}$^{+}$ & \meanstd{7.313}{0.079}$^{+}$ & 6.223$^{+}$ & 6.583$^{+}$ & \meanstd{7.143}{0.135}$^{+}$ & \meanstd{5.901}{0.097}$^{+}$ & \meanstdbf{5.628}{0.153}$^{-}$ & \meanstd{5.881}{0.105}$^{+}$ & \meanstd{5.695}{0.245}$^{=}$ & \meanstd{5.794}{0.045} \\
    & MAE & 4.485$^{+}$ & \meanstd{4.053}{0.013}$^{-}$ & \meanstd{4.148}{0.031}$^{=}$ & \meanstd{5.890}{0.094}$^{+}$ & 3.953$^{-}$ & 4.376$^{+}$ & \meanstd{4.877}{0.060}$^{+}$ & \meanstd{4.200}{0.057}$^{=}$ & \meanstd{3.895}{0.105}$^{-}$ & \meanstd{4.166}{0.093}$^{=}$ & \meanstdbf{3.873}{0.130}$^{-}$ & \meanstd{4.160}{0.053} \\
    \multirow{3}{*}{I3} & $R^2$ & 0.828$^{+}$ & \meanstdbf{0.878}{0.001}$^{=}$ & \textbf{\meanstd{0.878}{0.001}}$^{=}$ & \meanstd{0.741}{0.004}$^{+}$ & 0.854$^{+}$ & 0.773$^{+}$ & \meanstd{0.693}{0.010}$^{+}$ & \meanstd{0.648}{0.015}$^{+}$ & \meanstd{0.831}{0.005}$^{+}$ & \meanstd{0.853}{0.007}$^{+}$ & \meanstd{0.864}{0.004}$^{=}$ & \meanstd{0.866}{0.024} \\
    & RMSE & 4.990$^{+}$ & \meanstdbf{4.223}{0.019}$^{=}$ & \meanstd{4.226}{0.026}$^{=}$ & \meanstd{6.132}{0.047}$^{+}$ & 4.611$^{+}$ & 5.733$^{+}$ & \meanstd{6.645}{0.108}$^{+}$ & \meanstd{7.156}{0.159}$^{+}$ & \meanstd{4.965}{0.079}$^{+}$ & \meanstd{4.620}{0.107}$^{+}$ & \meanstd{4.442}{0.067}$^{=}$ & \meanstd{4.416}{0.374} \\
    & MAE & 3.628$^{+}$ & \meanstd{2.801}{0.013}$^{-}$ & \meanstd{2.807}{0.004}$^{-}$ & \meanstd{5.127}{0.050}$^{+}$ & \textbf{2.800}$^{-}$ & 3.763$^{+}$ & \meanstd{4.864}{0.115}$^{+}$ & \meanstd{5.081}{0.147}$^{+}$ & \meanstd{3.402}{0.059}$^{+}$ & \meanstd{3.292}{0.090}$^{+}$ & \meanstd{2.847}{0.047}$^{-}$ & \meanstd{3.102}{0.279} \\
    \multirow{3}{*}{I4} & $R^2$ & 0.757$^{+}$ & \meanstd{0.858}{0.001}$^{=}$ & \meanstd{0.848}{0.006}$^{+}$ & \meanstd{0.719}{0.006}$^{+}$ & 0.832$^{+}$ & 0.774$^{+}$ & \meanstd{0.720}{0.009}$^{+}$ & \meanstd{0.834}{0.006}$^{+}$ & \meanstd{0.847}{0.006}$^{+}$ & \meanstd{0.799}{0.015}$^{+}$ & \meanstd{0.831}{0.008}$^{+}$ & \meanstdbf{0.861}{0.023} \\
    & RMSE & 2.635$^{+}$ & \meanstdbf{2.046}{0.009}$^{=}$ & \meanstd{2.159}{0.020}$^{+}$ & \meanstd{2.948}{0.024}$^{+}$ & 2.336$^{+}$ & 2.640$^{+}$ & \meanstd{3.007}{0.041}$^{+}$ & \meanstd{2.315}{0.048}$^{+}$ & \meanstd{2.145}{0.047}$^{+}$ & \meanstd{2.206}{0.045}$^{+}$ & \meanstd{2.279}{0.069}$^{+}$ & \meanstd{2.070}{0.172} \\
    & MAE & 1.719$^{+}$ & \meanstdbf{1.275}{0.007}$^{=}$ & \meanstd{1.304}{0.011}$^{=}$ & \meanstd{2.262}{0.027}$^{+}$ & 1.277$^{=}$ & 1.683$^{+}$ & \meanstd{2.026}{0.034}$^{+}$ & \meanstd{1.561}{0.037}$^{+}$ & \meanstd{1.408}{0.030}$^{+}$ & \meanstd{1.458}{0.043}$^{+}$ & \meanstd{1.495}{0.038}$^{+}$ & \meanstd{1.289}{0.112} \\
    \multirow{3}{*}{I5} & $R^2$ & 0.688$^{+}$ & \meanstd{0.838}{0.001}$^{+}$ & \meanstd{0.827}{0.003}$^{+}$ & \meanstd{0.631}{0.007}$^{+}$ & 0.770$^{+}$ & 0.653$^{+}$ & \meanstd{0.624}{0.007}$^{+}$ & \meanstd{0.698}{0.005}$^{+}$ & \meanstd{0.795}{0.006}$^{+}$ & \meanstd{0.796}{0.011}$^{+}$ & \meanstd{0.806}{0.010}$^{+}$ & \meanstdbf{0.872}{0.008} \\
    & RMSE & 6.053$^{+}$ & \meanstd{4.420}{0.018}$^{+}$ & \meanstd{4.553}{0.050}$^{+}$ & \meanstd{6.785}{0.068}$^{+}$ & 5.160$^{+}$ & 6.260$^{+}$ & \meanstd{6.820}{0.065}$^{+}$ & \meanstd{6.166}{0.061}$^{+}$ & \meanstd{5.000}{0.081}$^{+}$ & \meanstd{4.902}{0.127}$^{+}$ & \meanstd{4.698}{0.108}$^{+}$ & \meanstdbf{3.902}{0.143} \\
    & MAE & 4.493$^{+}$ & \meanstd{2.829}{0.010}$^{+}$ & \meanstd{2.908}{0.039}$^{+}$ & \meanstd{5.693}{0.069}$^{+}$ & 3.095$^{+}$ & 3.979$^{+}$ & \meanstd{5.043}{0.083}$^{+}$ & \meanstd{4.408}{0.058}$^{+}$ & \meanstd{3.331}{0.070}$^{+}$ & \meanstd{3.475}{0.126}$^{+}$ & \meanstd{2.962}{0.062}$^{+}$ & \meanstdbf{2.651}{0.144} \\
    \multirow{3}{*}{I6} & $R^2$ & 0.747$^{+}$ & \meanstdbf{0.875}{0.002}$^{-}$ & \meanstd{0.855}{0.002}$^{-}$ & \meanstd{0.712}{0.012}$^{+}$ & 0.819$^{=}$ & 0.760$^{+}$ & \meanstd{0.817}{0.006}$^{=}$ & \meanstd{0.838}{0.007}$^{-}$ & \meanstd{0.848}{0.006}$^{-}$ & \meanstd{0.801}{0.013}$^{+}$ & \meanstd{0.831}{0.008}$^{=}$ & \meanstd{0.839}{0.037} \\
    & RMSE & 2.542$^{=}$ & \meanstdbf{2.032}{0.017}$^{-}$ & \meanstd{2.204}{0.019}$^{-}$ & \meanstd{2.723}{0.033}$^{+}$ & 2.450$^{-}$ & 2.480$^{-}$ & \meanstd{2.423}{0.037}$^{-}$ & \meanstd{2.401}{0.051}$^{-}$ & \meanstd{2.328}{0.047}$^{-}$ & \meanstd{2.338}{0.044}$^{-}$ & \meanstd{2.327}{0.045}$^{-}$ & \meanstd{2.605}{0.221} \\
    & MAE & 1.682$^{=}$ & \meanstdbf{1.210}{0.009}$^{-}$ & \meanstd{1.306}{0.007}$^{-}$ & \meanstd{2.141}{0.039}$^{+}$ & 1.409$^{-}$ & 1.545$^{-}$ & \meanstd{1.624}{0.028}$^{=}$ & \meanstd{1.495}{0.040}$^{-}$ & \meanstd{1.471}{0.029}$^{-}$ & \meanstd{1.519}{0.042}$^{-}$ & \meanstd{1.418}{0.025}$^{-}$ & \meanstd{1.660}{0.180} \\
    \midrule
    \multicolumn{2}{l}{Average Rank} & 8.111 & \textbf{2.750} & 4.028 & 11.389 & 5.778 & 9.500 & 10.722 & 7.944 & 4.500 & 6.167 & 3.778 & 3.333 \\
    \multicolumn{2}{l}{$+/=/-$} & 16/2/0 & 8/5/5 & 10/4/4 & 18/0/0 & 12/2/4 & 16/0/2 & 15/2/1 & 14/1/3 & 12/0/6 & 14/2/2 & 9/4/5 & -- \\
    \bottomrule
  \end{tabular}}
\replytableend

\textcolor{red}{Table~\ref{tab:analytical_cosydelay} shows that CoSyDelay achieves the best results at all six intersections and on all three metrics, with an average rank of 1.00. The Wilcoxon signed-rank tests on the paired differences further show that it significantly outperforms every standard and data-calibrated analytical model in all 18 intersection--metric comparisons.}

\textcolor{red}{Table~\ref{tab:datadriven-subset12-manuscript} shows that CoSyDelay remains competitive with flexible data-driven methods. It has the second-best average rank (3.33) among the 12 methods, achieves the best results on all three metrics at I1 and I5, and obtains the highest $R^2$ at I4. The paired tests show that CoSyDelay is significantly better than XGBoost in 10 of the 18 intersection--metric comparisons, statistically indistinguishable in four, and significantly worse in four. It also significantly outperforms AdaBoost in all 18 comparisons and linear regression, KNN, and GCN in 16, 16, and 15 comparisons, respectively. CoSyDelay therefore maintains competitive estimation accuracy while providing an explicit delay equation evaluated against traffic-delay principles.}

\begin{figure*}[!t]
\centering
\includegraphics[width=\textwidth]{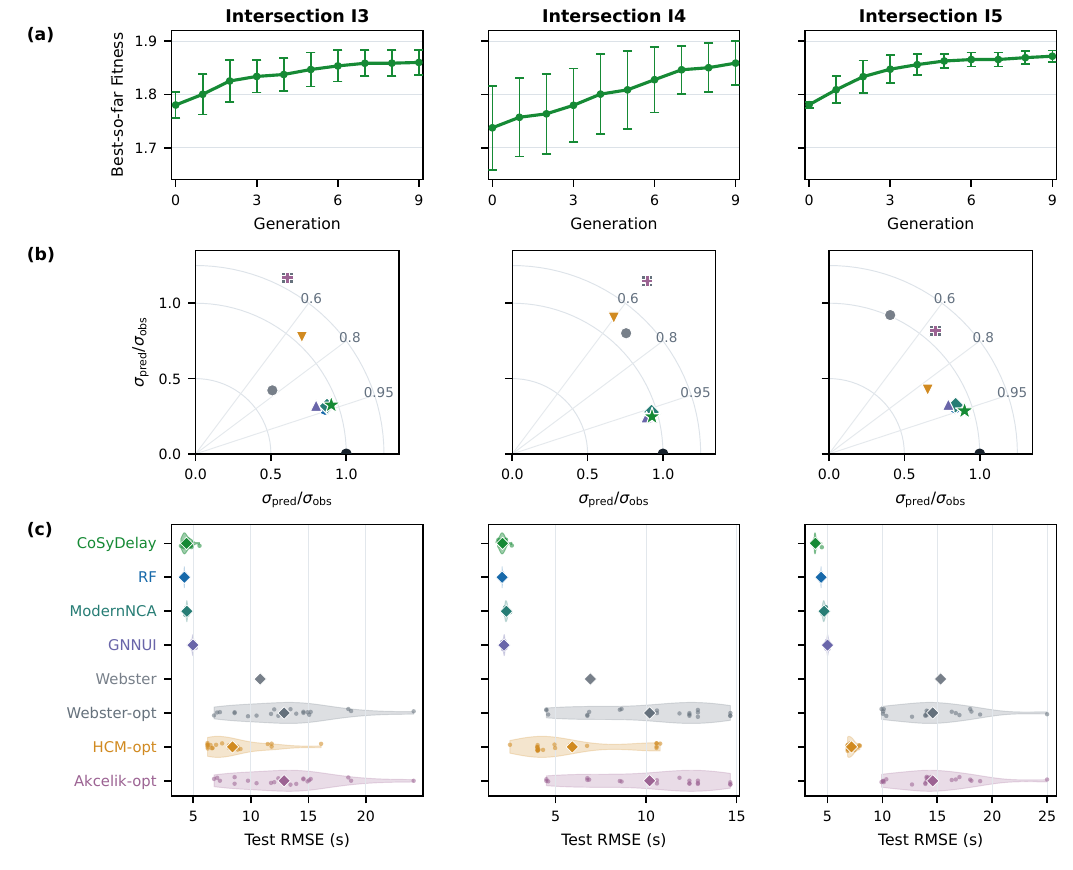}
\caption{\textcolor{red}{Visual analysis at I3--I5: (a) generation-wise best-so-far training Fitness, reported as the mean with capped $\pm$ sample-standard-deviation error bars over 20 independent runs; (b) test-set Taylor diagrams, with observations shown in black; and (c) test-RMSE violin plots, with diamonds marking the means. Method colors are shared by (b) and (c).}}
\label{fig:combined_visual_i3_i5}
\end{figure*}

\textcolor{red}{Fig.~\ref{fig:combined_visual_i3_i5} provides three complementary views of performance at I3--I5. First, the best-so-far Fitness increases throughout the search and begins to stabilize in the later generations; I4 shows the largest run-to-run variation. Fitness ranges from 0 to 2 and combines estimation accuracy with physical consistency. The final mean Fitness above 1.8 indicates strong combined performance in estimation accuracy and physical consistency. Second, the CoSyDelay points in the Taylor diagrams have correlations above 0.95 and normalized standard deviations close to 1, indicating close agreement with the observations and performance comparable to the strongest data-driven baselines. Third, its test-RMSE distributions are concentrated among the lowest values at all three intersections and remain substantially below those of the analytical models.}

\begin{figure}[!htbp]
        \centering
\includegraphics[width=\columnwidth]{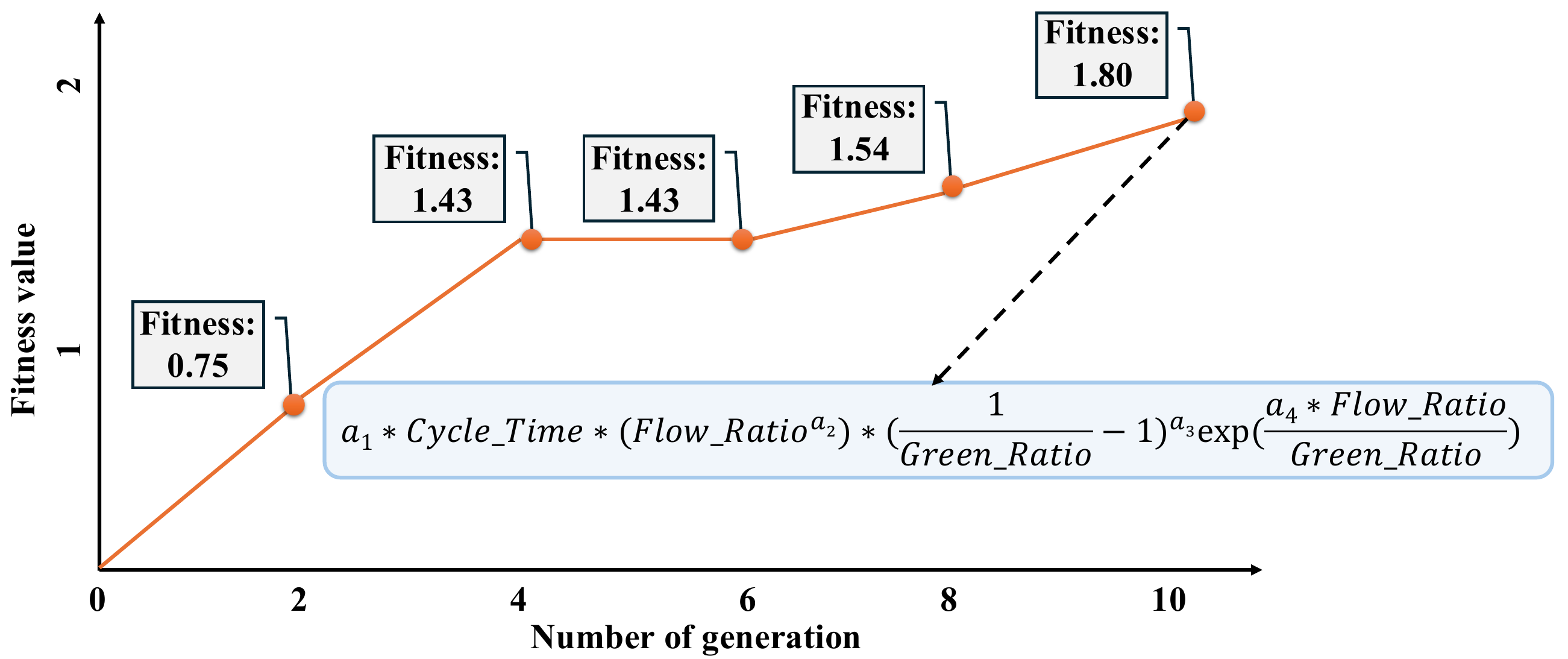}
\caption{
Evolution of the delay-expression structure at Intersection 1. Movement-specific coefficients are jointly calibrated against approach-level delay labels using L-BFGS-B, subject to $a_1,a_2,a_3,a_4>0$. The population size is 20 and the search runs for 10 generations.
}
        \label{fig:Evo_CoSydelay}
\par\vspace{8pt}
\centering
\captionof{table}{Estimated movement-specific parameters for turning movements at Intersection 1}
\label{tab:lane_parameters_improved}
\begingroup\footnotesize
\renewcommand{\arraystretch}{1.1}
\begin{tabular}{llcccc}

\toprule
\textbf{Approach} & \textbf{Movement} & $a_1$ & $a_2$ & $a_3$ & $a_4$ \\
\midrule
\multirow{3}{*}{South} & Left & 0.5700 & 0.2472 & 0.5925 & 0.3908 \\
                       & Through & 0.2967 & 0.4631 & 0.5595 & 0.0834 \\
                       & Right & 0.2823 & 0.4406 & 0.5095 & 0.4139 \\
\midrule
\multirow{3}{*}{East}  & Left & 1.0348 & 0.1901 & 0.0040 & 0.4159 \\
                       & Through & 0.1671 & 0.0010 & 0.3745 & 0.5062 \\
                       & Right & 0.7911 & 0.1872 & 0.0010 & 0.1121 \\
\midrule
\multirow{3}{*}{North} & Left & 0.3927 & 0.0119 & 0.3723 & 0.5713 \\
                       & Through & 0.2645 & 0.4595 & 0.5822 & 0.1339 \\
                       & Right & 0.4312 & 0.4513 & 0.3701 & 0.2235 \\
\midrule
\multirow{3}{*}{West}  & Left & 0.8289 & 0.1311 & 0.0010 & 1.1199 \\
                       & Through & 0.1670 & 0.0010 & 0.3306 & 0.5974 \\
                       & Right & 0.8570 & 0.1887 & 0.0010 & 0.0300 \\
\bottomrule

\end{tabular}
\endgroup
\par\vspace{8pt}
\captionof{table}{Performance changes under coefficient-sharing strategies at Intersection 1}
\label{tab:lane_parameters_sharing}
\begingroup\footnotesize
\renewcommand{\arraystretch}{1.15}
\begin{tabular}{lccc}
\toprule
\textbf{Strategy} & \textbf{$\Delta R^2$} & \textbf{$\Delta$RMSE} & \textbf{$\Delta$MAE} \\
\midrule
S$\rightarrow$N \;/\; E$\rightarrow$W
& $\downarrow 0.24\%$
& $\uparrow 0.43\%$
& $\downarrow 0.72\%$ \\
N$\rightarrow$S \;/\; W$\rightarrow$E
& $\downarrow 0.17\%$
& $\downarrow 0.28\%$
& $\uparrow 0.44\%$ \\
\bottomrule
\end{tabular}
\endgroup
\end{figure}

To gain deeper insight into the behavior of \textbf{CoSyDelay}, Fig.~\ref{fig:Evo_CoSydelay} shows the generation-wise trajectories of the learned models at Intersection~1. The best-so-far Fitness increases rapidly during the early generations and then stabilizes, indicating that the search progressively improves candidate expression structures and their calibrated coefficients. The selected expression for Intersection~1 is written as follows:
\begin{equation}
\label{eq:cosydelay}
\hat d  = a_1 \cdot C \cdot y^{a_2} \cdot \left(\frac{1}{\lambda} - 1\right)^{a_3} \cdot \exp\left(a_4 \cdot \frac{y}{\lambda}\right)
\end{equation}

\noindent where $C > 0$ is cycle time, $y$ is the flow ratio, and $\lambda$ is the green ratio. The resulting empirical model is parameterized with movement-specific coefficients $a_1, a_2, a_3, a_4 > 0$, jointly calibrated against approach-level delay labels and reported in Table~\ref{tab:lane_parameters_improved}. \textcolor{red}{Equation~\ref{eq:cosydelay} is the selected expression structure for Intersection~1; each final movement-level expression is obtained by substituting the corresponding calibrated coefficients. The expression is mathematically valid for \(y\geq0\), \(0<\lambda\leq1\), and \(C>0\). The condition \(\lambda>0\) avoids division by zero, while \(\lambda\leq1\) follows from the definition of the green ratio.}

We interpret the expression's factors, examine its mathematical behavior, and assess its movement-specific coefficients.

\begin{itemize}

    \item \textcolor{red}{\textbf{Physical Interpretation of the Discovered Expression:} The factor $a_1C$ sets the delay scale in seconds, while the remaining factors are dimensionless. The demand term $y^{a_2}$ describes the response to traffic flow, with $a_2$ controlling its power-law growth. The factor $(1/\lambda-1)^{a_3}$ describes the effect of the effective non-green-to-green time ratio: less effective green time increases this factor. Finally, $\exp(a_4y/\lambda)$ represents nonlinear growth with the degree of saturation, $X=y/\lambda$. These are interpretable response factors in an empirical formula, not independently identified delay components or rediscovered physical principles.}

% Instead of adopting classical rational delay formulations such as the Webster model or its variants---which rely on polynomial or rational terms like $\frac{X}{1-X}$ to approximate delay growth---the expression employ an exponential structure $\exp(a_4\cdot X)$. This exponential form captures the rapid, highly nonlinear escalation of delay in the vicinity of and beyond saturation ($X \ge 1$) more faithfully than traditional models.

        \item \textbf{Monotonicity and Boundary Behavior:} 
The evolved expression is monotonic in traffic state variables and satisfies delay boundary conditions:

	\begin{itemize}
	    \item \textit{Monotonicity:} 
	    In the interior domain \(y>0\) and \(0<\lambda<1\), the estimated delay \(\hat d\) increases strictly with the flow ratio (\(\partial \hat d/\partial y>0\)) and decreases strictly as the green ratio increases (\(\partial \hat d/\partial\lambda<0\)).

	    \item \textit{Compliance with Physical Boundaries:} 
	    The formulation satisfies fundamental boundary conditions at extreme operational limits:
    \begin{enumerate}
        \item \textcolor{red}{\textbf{Low-Flow Condition ($y \to 0^+$):} As traffic demand vanishes, the term $y^{a_2}$ approaches zero (given $a_2 > 0$), resulting in $\hat d \to 0$. This is a finite, non-negative low-flow limit. Zero is a valid special case; other per-vehicle delay expressions may approach a positive finite limit because an isolated vehicle can still arrive during a red interval.}
        \item \textbf{Free-Flow Condition ($\lambda \to 1$):} As the green ratio approaches unity (continuous green), the geometric impedance term $(1/\lambda - 1)$ becomes zero, leading to $\hat d \to 0$. This correctly models free-flow movement where signalized interruption is absent.
        \item \textcolor{red}{\textbf{Zero Green Limit ($\lambda \to 0^+$ with $y>0$):} Under sustained positive demand, both the red-to-green term and the exponential saturation term diverge as the green ratio approaches zero, yielding $\hat d\to+\infty$. This represents the loss of service when no green time is allocated.}
    \end{enumerate}
\end{itemize}

    \item \textbf{Movement-Specific Coefficients:} The coefficients allow the shared structure to represent different turning-movement responses: $a_1$ controls the delay scale, while $a_2$, $a_3$, and $a_4$ adjust sensitivity to demand, green allocation, and saturation, respectively. We tested coefficient sharing between opposing movements at Intersection~1. Table~\ref{tab:lane_parameters_sharing} shows that all metric changes remain below $0.8\%$, indicating limited sensitivity to sharing in this tested configuration. This result does not imply general coefficient interchangeability across different geometries or signal plans; movement-specific coefficients are therefore jointly calibrated against approach-level delay labels.

\end{itemize}

\subsection{\textcolor{red}{Generalization Analysis}}
\label{sec:generalization_analysis}
\textcolor{red}{We evaluate CoSyDelay's generalization ability under oversaturated demand and across intersections, and further assess its utility for offline signal-timing optimization at new simulated intersections.}

\subsubsection{\textcolor{red}{Delay-Estimation Generalization}}
\textcolor{red}{Delay-estimation generalization is evaluated in two settings. First, each intersection's learned expression and coefficients are fixed and evaluated on 3,644 additional simulated samples under oversaturated conditions (degree of saturation \(X=y/\lambda\geq1.00\)). These samples are not used for model development. As shown in Table~\ref{tab:train_test_comparison_no_fitness}, CoSyDelay ranks first, with mean $R^2=0.841$, RMSE $=3.754$~s, and MAE $=2.630$~s, showing that it maintains strong estimation performance under the tested oversaturated regime. Second, cross-intersection structure transfer is evaluated over all 30 ordered source--target pairs among I1--I6. For each pair, the expression structure discovered at the source intersection is transferred unchanged, its coefficients are recalibrated using the target-intersection training data, and the resulting model is evaluated on the target test set. CoSyDelay again ranks first, attaining mean $R^2=0.856$, RMSE $=4.477$~s, and MAE $=2.937$~s. These results show that the discovered expression structures remain effective after target-specific coefficient calibration.}

\begin{table*}[!h]
\centering
\color{red}
\caption{Delay-estimation generalization under oversaturated demand and cross-intersection structure transfer. Entries are mean $\pm$ standard deviation across six intersection-level results for oversaturated traffic (3,644 samples in total) and across 30 ordered source--target structure transfers. Average Rank combines the $R^2$, RMSE, and MAE ranks within each experiment (lower is better). Best results are in bold.}
\label{tab:train_test_comparison_no_fitness}
\footnotesize
\setlength{\tabcolsep}{1pt}
\renewcommand{\arraystretch}{1.15}
\begin{tabular*}{\textwidth}{@{\extracolsep{\fill}}lcccccccc@{}}
\toprule
\multirow{2}{*}{\textbf{Model}} & \multicolumn{4}{c}{\textbf{Oversaturated traffic}} & \multicolumn{4}{c}{\textbf{Cross-intersection structure transfer}} \\
\cmidrule(lr){2-5}\cmidrule(lr){6-9}
& $R^2$ & RMSE (s) & MAE (s) & Average Rank & $R^2$ & RMSE (s) & MAE (s) & Average Rank \\
\midrule
Webster & \meanstd{-57.232}{102.708} & \meanstd{15.021}{5.201} & \meanstd{9.009}{2.540} & 6.00 & \meanstd{-0.003}{0.340} & \meanstd{11.881}{3.401} & \meanstd{8.161}{3.104} & 6.44 \\
Webster-optimized & \meanstd{-0.008}{0.924} & \meanstd{6.411}{1.442} & \meanstd{4.663}{1.171} & 3.94 & \meanstd{0.606}{0.228} & \meanstd{7.241}{2.477} & \meanstd{4.538}{0.845} & 4.17 \\
HCM & \meanstd{-0.025}{0.324} & \meanstd{9.685}{4.140} & \meanstd{7.821}{3.447} & 5.11 & \meanstd{0.259}{0.380} & \meanstd{10.184}{4.101} & \meanstd{7.843}{3.443} & 5.25 \\
HCM-optimized & \meanstd{0.265}{0.851} & \meanstd{5.411}{1.322} & \meanstd{3.937}{1.038} & 2.89 & \meanstd{0.790}{0.078} & \meanstd{5.477}{1.817} & \meanstd{3.565}{1.538} & 2.10 \\
Ak\c{c}elik & \meanstd{-0.025}{0.324} & \meanstd{9.685}{4.140} & \meanstd{7.821}{3.447} & 5.11 & \meanstd{0.259}{0.380} & \meanstd{10.184}{4.101} & \meanstd{7.843}{3.443} & 5.25 \\
Ak\c{c}elik-optimized & \meanstd{-0.008}{0.924} & \meanstd{6.411}{1.442} & \meanstd{4.663}{1.171} & 3.94 & \meanstd{0.630}{0.231} & \meanstd{7.047}{2.589} & \meanstd{4.293}{0.841} & 3.56 \\
\textbf{CoSyDelay} & \meanstdbf{0.841}{0.052} & \meanstdbf{3.754}{1.301} & \meanstdbf{2.630}{1.036} & \textbf{1.00} & \meanstdbf{0.856}{0.038} & \meanstdbf{4.477}{1.122} & \meanstdbf{2.937}{0.916} & \textbf{1.23} \\
\bottomrule
\end{tabular*}

\end{table*}

\subsubsection{\textcolor{red}{Downstream Signal-Timing Optimization}}
\textcolor{red}{We further test whether the fixed CoSyDelay model learned at I1 can support offline signal-timing decisions at new intersections. Its expression and coefficients are applied unchanged to three newly constructed intersection scenarios in Simulation of Urban MObility (SUMO) \cite{behrisch2011sumo}: Beijing Gaojiaoyuan, Chengdu Guanghua, and Tianjin Zhijingdao. Each scenario is tested under five traffic regimes. For each intersection--regime combination, each delay model selects the candidate timing plan with the lowest estimated delay, and the selected plan is simulated using 20 random seeds. CoSyDelay is compared with the standard and optimized variants of Webster, HCM, and Ak\c{c}elik; none of the models is calibrated on the three new scenarios. Table~\ref{tab:downstream_signal_optimization} shows that CoSyDelay produces the lowest delay at all three intersections. Its overall mean delay is 93.753~s, compared with 119.507~s for the standard analytical models and 172.195~s for their optimized variants, corresponding to reductions of 21.5\% and 45.6\%, respectively. These results demonstrate that CoSyDelay can support effective offline timing-plan selection in the tested scenarios. }

\begin{table}[!t]
\centering
\color{red}
\caption{\textcolor{red}{Signal-timing optimization at three new simulated intersections. Values are mean $\pm$ standard deviation of average total delay per vehicle (s); lower is better.}}
\label{tab:downstream_signal_optimization}
\scriptsize
\setlength{\tabcolsep}{2pt}
\renewcommand{\arraystretch}{1.08}
\begin{tabular*}{\columnwidth}{@{\extracolsep{\fill}}lccc@{}}
\toprule
\textbf{Model} & \textbf{Beijing} & \textbf{Chengdu} & \textbf{Tianjin} \\
\midrule
Webster & \meanstd{120.271}{48.377} & \meanstd{124.741}{50.762} & \meanstd{113.508}{48.902} \\
HCM & \meanstd{120.271}{48.377} & \meanstd{124.741}{50.762} & \meanstd{113.508}{48.902} \\
Ak\c{c}elik & \meanstd{120.271}{48.377} & \meanstd{124.741}{50.762} & \meanstd{113.508}{48.902} \\
Webster-optimized & \meanstd{172.262}{61.485} & \meanstd{182.415}{77.177} & \meanstd{161.908}{75.922} \\
HCM-optimized & \meanstd{172.262}{61.485} & \meanstd{182.415}{77.177} & \meanstd{161.908}{75.922} \\
Ak\c{c}elik-optimized & \meanstd{172.262}{61.485} & \meanstd{182.415}{77.177} & \meanstd{161.908}{75.922} \\
\textbf{CoSyDelay} & \meanstdbf{92.943}{43.852} & \meanstdbf{93.737}{41.348} & \meanstdbf{94.578}{50.433} \\
\bottomrule
\end{tabular*}
\end{table}

% Wide table: use the separate double-column float queue to avoid blocking single-column tables; captionof{table} preserves table numbering.
\begin{figure*}[!b]
\centering
\captionof{table}{\textcolor{red}{Component ablation and symbolic-search comparison at I1--I6. Values are test Fitness, reported as mean $\pm$ standard deviation over 20 runs. Average Rank is calculated within each block (lower is better); $+/=/-$ summarizes Holm-adjusted Wilcoxon signed-rank tests on the paired differences relative to CoSyDelay. Symbolic-search methods use a common 600-s budget. Best values are bold.}}
\label{tab:ablation_fitness_comparison}
\label{tab:binary_principlewise_fitness}
\label{tab:search_fitness_comparison}
\label{tab:controlled_structure_search}
\scriptsize
\setlength{\tabcolsep}{2pt}
\renewcommand{\arraystretch}{1.12}
\newcommand{\ablationredrow}[9]{\textcolor{red}{#1} & \textcolor{red}{#2} & \textcolor{red}{#3} & \textcolor{red}{#4} & \textcolor{red}{#5} & \textcolor{red}{#6} & \textcolor{red}{#7} & \textcolor{red}{#8} & \textcolor{red}{#9} \\}
\begin{tabular*}{\textwidth}{@{\extracolsep{\fill}}lcccccccc@{}}
\toprule
\textbf{Method} & \textbf{I1} & \textbf{I2} & \textbf{I3} & \textbf{I4} & \textbf{I5} & \textbf{I6} & \textbf{Average Rank} & \textbf{$+/=/-$} \\
\midrule
\multicolumn{9}{l}{\textit{Input feature representation}} \\
\cmidrule(lr){1-9}
Raw inputs & \meanstd{1.717}{0.001} & \meanstd{1.656}{0.001} & \meanstd{1.571}{0.002} & \meanstd{1.530}{0.008} & \meanstd{1.619}{0.003} & \meanstd{1.733}{0.003} & 2.000 & 6/0/0 \\
Transformed features (CoSyDelay) & \meanstdbf{1.803}{0.002} & \meanstdbf{1.756}{0.003} & \meanstdbf{1.866}{0.024} & \meanstdbf{1.861}{0.023} & \meanstdbf{1.872}{0.008} & \meanstdbf{1.839}{0.037} & \textbf{1.000} & -- \\
\midrule
\multicolumn{9}{l}{\textit{Modeling granularity}} \\
\cmidrule(lr){1-9}
Intersection model & \meanstd{0.923}{0.001} & \meanstd{0.853}{0.004} & \meanstd{1.664}{0.003} & \meanstd{1.134}{0.549} & \meanstd{1.349}{0.546} & \meanstd{0.537}{0.450} & 2.167 & 6/0/0 \\
Approach model & \meanstd{0.715}{0.001} & \meanstd{0.653}{0.001} & \meanstd{0.558}{0.001} & \meanstd{0.311}{0.003} & \meanstd{0.460}{0.002} & \meanstd{0.551}{0.004} & 2.833 & 6/0/0 \\
Turning-movement model (CoSyDelay) & \meanstdbf{1.803}{0.002} & \meanstdbf{1.756}{0.003} & \meanstdbf{1.866}{0.024} & \meanstdbf{1.861}{0.023} & \meanstdbf{1.872}{0.008} & \meanstdbf{1.839}{0.037} & \textbf{1.000} & -- \\
\midrule
\multicolumn{9}{l}{\textcolor{red}{\textit{Physical-consistency scoring}}} \\
\cmidrule(lr){1-9}
\ablationredrow{Principlewise score}{\meanstd{1.792}{0.003}}{\meanstdbf{1.757}{0.003}}{\meanstd{1.865}{0.024}}{\meanstd{1.860}{0.041}}{\meanstd{1.867}{0.011}}{\meanstd{1.826}{0.018}}{1.833}{0/6/0}
\ablationredrow{Binary score (CoSyDelay)}{\meanstdbf{1.803}{0.002}}{\meanstd{1.756}{0.003}}{\meanstdbf{1.866}{0.024}}{\meanstdbf{1.861}{0.023}}{\meanstdbf{1.872}{0.008}}{\meanstdbf{1.839}{0.037}}{\textbf{1.167}}{--}
\midrule
% TODO: The restored SR significance markers are from the previous comparison; recheck paired-run tests against the updated CoSyDelay results before submission.
\multicolumn{9}{l}{\textcolor{red}{\textit{Symbolic-search methods (600-s budget)}}} \\
\cmidrule(lr){1-9}
\ablationredrow{Operon}{\meanstd{0.829}{0.013}}{\meanstd{0.834}{0.003}}{\meanstd{0.916}{0.014}}{\meanstd{0.866}{0.014}}{\meanstd{0.902}{0.005}}{\meanstd{0.886}{0.020}}{2.167}{6/0/0}
\ablationredrow{PySR}{\meanstd{0.827}{0.006}}{\meanstd{0.745}{0.255}}{\meanstd{0.914}{0.010}}{\meanstd{0.863}{0.018}}{\meanstd{0.898}{0.006}}{\meanstd{0.894}{0.004}}{3.000}{6/0/0}
\ablationredrow{GP-GOMEA}{\meanstd{0.767}{0.027}}{\meanstd{0.756}{0.031}}{\meanstd{0.779}{0.038}}{\meanstd{0.810}{0.042}}{\meanstd{0.856}{0.026}}{\meanstd{0.872}{0.043}}{3.833}{6/0/0}
\ablationredrow{CoSyDelay}{\meanstdbf{1.792}{0.002}}{\meanstdbf{1.752}{0.006}}{\meanstdbf{1.869}{0.033}}{\meanstdbf{1.858}{0.037}}{\meanstdbf{1.844}{0.064}}{\meanstdbf{1.832}{0.005}}{\textbf{1.000}}{--}
\midrule
\multicolumn{9}{l}{\textcolor{revisionRtwo}{\textit{Controlled structure search (600-s budget)}}} \\
\cmidrule(lr){1-9}
\textcolor{revisionRtwo}{GA} & \textcolor{revisionRtwo}{\meanstd{1.787}{0.004}} & \textcolor{revisionRtwo}{\meanstd{1.750}{0.006}} & \textcolor{revisionRtwo}{\meanstd{1.792}{0.052}} & \textcolor{revisionRtwo}{\meanstd{1.786}{0.047}} & \textcolor{revisionRtwo}{\meanstd{1.798}{0.019}} & \textcolor{revisionRtwo}{\meanstd{1.761}{0.088}} & \textcolor{revisionRtwo}{2.000} & \textcolor{revisionRtwo}{6/0/0} \\
\textcolor{revisionRtwo}{Random search} & \textcolor{revisionRtwo}{\meanstd{1.731}{0.004}} & \textcolor{revisionRtwo}{\meanstd{1.709}{0.019}} & \textcolor{revisionRtwo}{\meanstd{1.492}{0.521}} & \textcolor{revisionRtwo}{\meanstd{1.391}{0.213}} & \textcolor{revisionRtwo}{\meanstd{1.637}{0.037}} & \textcolor{revisionRtwo}{\meanstd{0.854}{0.476}} & \textcolor{revisionRtwo}{3.333} & \textcolor{revisionRtwo}{6/0/0} \\
\textcolor{revisionRtwo}{MCTS} & \textcolor{revisionRtwo}{\meanstd{1.744}{0.024}} & \textcolor{revisionRtwo}{\meanstd{1.707}{0.016}} & \textcolor{revisionRtwo}{\meanstd{1.348}{0.357}} & \textcolor{revisionRtwo}{\meanstd{1.273}{0.390}} & \textcolor{revisionRtwo}{\meanstd{1.463}{0.393}} & \textcolor{revisionRtwo}{\meanstd{1.506}{0.221}} & \textcolor{revisionRtwo}{3.667} & \textcolor{revisionRtwo}{6/0/0} \\
\textcolor{revisionRtwo}{CoSyDelay} & \textcolor{revisionRtwo}{\meanstdbf{1.792}{0.002}} & \textcolor{revisionRtwo}{\meanstdbf{1.752}{0.006}} & \textcolor{revisionRtwo}{\meanstdbf{1.869}{0.033}} & \textcolor{revisionRtwo}{\meanstdbf{1.858}{0.037}} & \textcolor{revisionRtwo}{\meanstdbf{1.844}{0.064}} & \textcolor{revisionRtwo}{\meanstdbf{1.832}{0.005}} & \textcolor{revisionRtwo}{\textbf{1.000}} & \textcolor{revisionRtwo}{--} \\
\bottomrule
\end{tabular*}
\end{figure*}

\subsection{\textcolor{red}{Component Ablation Analysis}}
\label{sec:ablation_analysis}

\subsubsection{Input Feature Transformation}
We compare raw movement-level inputs with the proposed feature transformation under the same search and calibration procedure. CoSyDelay converts traffic volume and green time into the flow ratio and green ratio, respectively, while retaining cycle length. The transformed inputs describe demand relative to capacity and the effective green allocation, providing variables that are directly meaningful for delay estimation. Table~\ref{tab:ablation_fitness_comparison} shows that this transformation improves Fitness at all six intersections.

\subsubsection{Modeling Granularity}
We compare intersection-, approach-, and turning-movement-level formulations. CoSyDelay learns expressions at the turning-movement level and aggregates their estimates to the approach and intersection levels. Table~\ref{tab:ablation_fitness_comparison} shows that this formulation achieves the highest Fitness at every intersection, with an average rank of 1.00 versus 2.17 and 2.83 for the intersection- and approach-level formulations, respectively.

\subsubsection{\textcolor{red}{Physical-Consistency Scoring}}
\textcolor{red}{We isolate the effect of physical-consistency scoring by keeping the prompt, candidate-generation procedure, coefficient calibration, and search budget unchanged. The default binary score assigns $\mathrm{Score}_{\mathrm{phys}}=1$ only when all seven traffic-delay principles are satisfied and 0 otherwise. The principlewise alternative averages the seven individual pass indicators and therefore gives partial credit when only some principles are satisfied. Both schemes assign a score of 1 to fully compliant expressions; the principlewise scheme only distinguishes among partially compliant expressions. No statistically significant improvement in test Fitness is observed with principlewise scoring. We therefore retain the simpler binary score, which gives physical credit only when all required principles pass.}

\subsubsection{\textcolor{red}{Comparison with Symbolic-Search Methods}}
\textcolor{red}{In CoSyDelay, the LLM proposes and revises expression structures, while movement-specific calibration and traffic-principle evaluation complete the search process. We compare CoSyDelay with Operon~\cite{burlacu2020operon}, PySR~\cite{cranmer2023pysr}, the Gene-pool Optimal Mixing Evolutionary Algorithm for genetic programming (GP-GOMEA)~\cite{virgolin2021improving} under the same 600-s wall-clock budget.} \textcolor{revisionRtwo}{This total wall-clock budget includes API waiting time, failed calls and retries, and coefficient-calibration time.} \textcolor{red}{ CoSyDelay ranks first in test Fitness at all six intersections (average rank 1.00). Its selected expressions satisfy the prescribed traffic-delay principles ($\mathrm{Score}_{\mathrm{phys}}=1$), whereas all competing methods violate at least one prescribed traffic-delay principle ($\mathrm{Score}_{\mathrm{phys}}=0$). }

\textcolor{revisionRtwo}{In addition, we report estimation accuracy and physical compliance separately for CoSyDelay and the three competing SR methods in Table~\ref{tab:sr_accuracy_physical}. Operon and PySR achieve higher accuracy than CoSyDelay, but none of their runs produces an expression satisfying all seven principles simultaneously, whereas CoSyDelay satisfies all seven in every run. These results highlight a trade-off: satisfying physical requirements may entail some loss of estimation accuracy, while CoSyDelay aims to balance the two objectives.}

\textcolor{revisionRtwo}{To isolate the contribution of LLM-based structure search, we additionally compare CoSyDelay with a genetic algorithm (GA), random search, and Monte Carlo tree search (MCTS) within the same pipeline. Only the structure-search method changes; input features, movement decomposition, coefficient calibration, and physical evaluation are held fixed. Each method uses the same 600-s wall-clock budget and is evaluated over 20 runs at each of the six intersections. Table~\ref{tab:controlled_structure_search} reports test Fitness in the controlled-search block. CoSyDelay achieves the highest $R^2$ and lowest RMSE and MAE at every intersection, with six-intersection macro-averages of 0.8245, 3.940~s, and 2.723~s, respectively. These results support the contribution of LLM-based structure search to estimation accuracy within the fixed pipeline.}

\begin{table}[!t]
\centering
\color{revisionRtwo}
\caption{Accuracy and physical compliance under a 600-s budget (20 runs per intersection). Accuracy metrics are six-intersection macro-averages (mean $\pm$ SD). ``Yes'' indicates that all runs pass all seven principles; ``No'' indicates that no run passes all seven simultaneously. Best values are bold.}
\label{tab:sr_accuracy_physical}
\scriptsize
\setlength{\tabcolsep}{1.5pt}
\renewcommand{\arraystretch}{1.12}
\begin{tabular*}{\columnwidth}{@{\extracolsep{\fill}}lcccc@{}}
\toprule
Method & $R^2\uparrow$ & RMSE (s)$\downarrow$ & MAE (s)$\downarrow$ & \shortstack{Physical\\compliance} \\
\midrule
Operon & \meanstdbf{0.8721}{0.0049} & \meanstdbf{3.186}{0.071} & \meanstdbf{2.179}{0.018} & No \\
PySR & \meanstd{0.8567}{0.0448} & \meanstd{3.321}{0.295} & \meanstd{2.224}{0.035} & No \\
GP-GOMEA & \meanstd{0.8067}{0.0126} & \meanstd{4.075}{0.154} & \meanstd{2.854}{0.106} & No \\
CoSyDelay & \meanstd{0.8245}{0.0084} & \meanstd{3.940}{0.115} & \meanstd{2.723}{0.099} & \textbf{Yes} \\
\bottomrule
\end{tabular*}
\end{table}

% Wide table: use the separate double-column float queue to avoid blocking single-column tables; captionof{table} preserves table numbering.
\begin{figure*}[!b]
\centering
\captionof{table}{\textcolor{red}{Ablation and sensitivity analysis of LLM-guided search across six intersections. Values are test Fitness, reported as mean $\pm$ standard deviation over 20 runs. Average Rank is calculated within each block (lower is better); $+/=/-$ counts the intersections where CoSyDelay is significantly better, not significantly different, or significantly worse than each alternative. Best values are in bold.}}
\label{tab:llm_parameter_sensitivity}
\label{tab:parameter_fitness_comparison}
\label{tab:temperature_sensitivity}
\label{tab:prompt_evaluator_ablation}
\scriptsize
\setlength{\tabcolsep}{1.5pt}
\renewcommand{\arraystretch}{1.0}
\setlength{\aboverulesep}{1pt}
\setlength{\belowrulesep}{1pt}
\newcommand{\llmredrow}[9]{\textcolor{red}{#1} & \textcolor{red}{#2} & \textcolor{red}{#3} & \textcolor{red}{#4} & \textcolor{red}{#5} & \textcolor{red}{#6} & \textcolor{red}{#7} & \textcolor{red}{#8} & \textcolor{red}{#9} \\}
\begin{tabular*}{\textwidth}{@{\extracolsep{\fill}}lcccccccc@{}}
\toprule
\textbf{Setting} & \textbf{I1} & \textbf{I2} & \textbf{I3} & \textbf{I4} & \textbf{I5} & \textbf{I6} & \textbf{Average Rank} & \textbf{$+/=/-$} \\
\midrule
\multicolumn{9}{l}{\textit{LLM backend}} \\
\cmidrule(lr){1-9}
Qwen-Plus & \meanstd{1.790}{0.001} & \meanstd{1.741}{0.005} & \meanstd{1.841}{0.013} & \meanstd{1.658}{0.084} & \meanstd{1.862}{0.011} & \meanstd{1.763}{0.021} & 4.417 & 2/4/0 \\
Gemini-2.5-Flash & \meanstd{1.790}{0.000} & \meanstdbf{1.759}{0.001} & \meanstdbf{1.885}{0.004} & \meanstdbf{1.880}{0.012} & \meanstd{1.869}{0.006} & \meanstd{1.827}{0.001} & 2.000 & 0/6/0 \\
GPT-OSS-20B & \meanstd{1.793}{0.000} & \meanstd{1.755}{0.004} & \meanstdbf{1.885}{0.014} & \meanstd{1.839}{0.013} & \meanstd{1.796}{0.004} & \meanstd{1.813}{0.018} & 3.167 & 1/5/0 \\
Grok-Code-Fast-1 & \meanstd{1.779}{0.012} & \meanstd{1.744}{0.008} & \meanstd{1.847}{0.000} & \meanstd{1.840}{0.007} & \meanstd{1.796}{0.017} & \meanstd{1.830}{0.003} & 3.750 & 2/4/0 \\
GPT-4.1-mini (CoSyDelay) & \meanstdbf{1.803}{0.002} & \meanstd{1.756}{0.003} & \meanstd{1.866}{0.024} & \meanstd{1.861}{0.023} & \meanstdbf{1.872}{0.008} & \meanstdbf{1.839}{0.037} & \textbf{1.667} & -- \\
\midrule
\multicolumn{9}{l}{\textcolor{red}{\textit{Principle-informed prompt and evaluator}}} \\
\cmidrule(lr){1-9}
\llmredrow{Neither}{\meanstd{0.817}{0.004}}{\meanstd{0.769}{0.008}}{\meanstd{0.816}{0.018}}{\meanstd{0.842}{0.022}}{\meanstd{0.769}{0.030}}{\meanstd{0.836}{0.006}}{3.667}{6/0/0}
\llmredrow{Prompt only}{\meanstd{0.815}{0.001}}{\meanstd{0.772}{0.007}}{\meanstd{0.879}{0.009}}{\meanstd{0.862}{0.012}}{\meanstd{0.862}{0.002}}{\meanstd{0.796}{0.017}}{3.333}{6/0/0}
\llmredrow{Evaluator only}{\meanstd{1.448}{0.068}}{\meanstd{1.380}{0.039}}{\meanstd{1.415}{0.036}}{\meanstd{1.439}{0.033}}{\meanstd{1.357}{0.103}}{\meanstd{1.452}{0.045}}{2.000}{6/0/0}
\llmredrow{Both ON (CoSyDelay)}{\meanstdbf{1.803}{0.002}}{\meanstdbf{1.756}{0.003}}{\meanstdbf{1.866}{0.024}}{\meanstdbf{1.861}{0.023}}{\meanstdbf{1.872}{0.008}}{\meanstdbf{1.839}{0.037}}{\textbf{1.000}}{--}
\midrule
\multicolumn{9}{l}{\textcolor{red}{\textit{Principle instruction and temperature}}} \\
\cmidrule(lr){1-9}
\llmredrow{OFF, $T=0.0$}{\meanstd{0.789}{0.013}}{\meanstd{0.741}{0.017}}{\meanstd{0.839}{0.008}}{\meanstd{0.782}{0.050}}{\meanstd{0.814}{0.061}}{\meanstd{1.057}{0.413}}{5.167}{6/0/0}
\llmredrow{OFF, $T=0.3$}{\meanstd{0.802}{0.006}}{\meanstd{0.754}{0.023}}{\meanstd{0.795}{0.033}}{\meanstd{0.813}{0.017}}{\meanstd{0.812}{0.040}}{\meanstd{0.818}{0.008}}{5.167}{6/0/0}
\llmredrow{OFF, $T=0.7$}{\meanstd{0.807}{0.007}}{\meanstd{0.748}{0.030}}{\meanstd{0.822}{0.030}}{\meanstd{0.792}{0.014}}{\meanstd{0.821}{0.037}}{\meanstd{0.842}{0.032}}{4.667}{6/0/0}
\llmredrow{ON, $T=0.0$ (CoSyDelay)}{\meanstdbf{1.803}{0.002}}{\meanstd{1.756}{0.003}}{\meanstdbf{1.866}{0.024}}{\meanstdbf{1.861}{0.023}}{\meanstdbf{1.872}{0.008}}{\meanstdbf{1.839}{0.037}}{\textbf{1.333}}{--}
\llmredrow{ON, $T=0.3$}{\meanstd{1.800}{0.004}}{\meanstdbf{1.762}{0.004}}{\meanstd{1.765}{0.026}}{\meanstd{1.653}{0.055}}{\meanstd{1.749}{0.005}}{\meanstd{1.301}{0.422}}{2.333}{0/6/0}
\llmredrow{ON, $T=0.7$}{\meanstd{1.798}{0.007}}{\meanstd{1.758}{0.011}}{\meanstd{1.792}{0.043}}{\meanstd{1.648}{0.041}}{\meanstd{1.755}{0.022}}{\meanstd{1.388}{0.336}}{2.333}{0/6/0}
\midrule
\multicolumn{9}{l}{\textit{Generation--population setting $(G,P)$}} \\
\cmidrule(lr){1-9}
$(5,5)$ & \meanstd{1.791}{0.003} & \meanstd{1.746}{0.009} & \meanstd{1.878}{0.002} & \meanstd{1.805}{0.023} & \meanstd{1.870}{0.000} & \meanstd{1.756}{0.026} & 7.167 & 2/4/0 \\
$(5,10)$ & \meanstd{1.791}{0.001} & \meanstd{1.756}{0.004} & \meanstd{1.815}{0.065} & \meanstd{1.866}{0.025} & \meanstd{1.865}{0.009} & \meanstd{1.836}{0.000} & 6.417 & 1/5/0 \\
$(5,20)$ & \meanstd{1.788}{0.000} & \meanstd{1.757}{0.002} & \meanstd{1.879}{0.001} & \meanstd{1.851}{0.011} & \meanstd{1.857}{0.017} & \meanstd{1.833}{0.001} & 6.833 & 0/6/0 \\
$(10,5)$ & \meanstd{1.793}{0.000} & \meanstd{1.757}{0.002} & \meanstd{1.854}{0.026} & \meanstd{1.853}{0.006} & \meanstd{1.870}{0.000} & \meanstd{1.784}{0.052} & 6.083 & 1/5/0 \\
$(10,10)$ & \meanstd{1.790}{0.002} & \meanstd{1.758}{0.001} & \meanstd{1.863}{0.017} & \meanstd{1.890}{0.000} & \meanstd{1.872}{0.002} & \meanstd{1.834}{0.004} & 4.750 & 0/6/0 \\
$(10,20)$ (CoSyDelay) & \meanstdbf{1.803}{0.002} & \meanstd{1.756}{0.003} & \meanstd{1.866}{0.024} & \meanstd{1.861}{0.023} & \meanstd{1.872}{0.008} & \meanstd{1.839}{0.037} & 4.417 & -- \\
$(20,5)$ & \meanstd{1.789}{0.001} & \meanstd{1.757}{0.002} & \meanstd{1.880}{0.000} & \meanstd{1.890}{0.000} & \meanstd{1.862}{0.009} & \meanstd{1.839}{0.000} & 4.833 & 0/6/0 \\
$(20,10)$ & \meanstd{1.795}{0.003} & \meanstd{1.756}{0.003} & \meanstdbf{1.896}{0.001} & \meanstdbf{1.892}{0.002} & \meanstd{1.876}{0.002} & \meanstd{1.839}{0.000} & 2.833 & 0/6/0 \\
$(20,20)$ & \meanstd{1.797}{0.002} & \meanstdbf{1.761}{0.002} & \meanstd{1.883}{0.013} & \meanstd{1.890}{0.000} & \meanstdbf{1.877}{0.000} & \meanstdbf{1.847}{0.007} & \textbf{1.667} & 0/6/0 \\
\bottomrule
\end{tabular*}
\end{figure*}

\subsection{\textcolor{red}{Effectiveness and Sensitivity of LLM-Guided Search}}
\label{sec:llm_effectiveness}
\label{sec:parameter_analysis}
\textcolor{red}{We assess the LLM-guided search by varying four factors: the LLM backend, the principle-informed prompt and evaluator, generation temperature, and the generation--population setting. Each setting is evaluated over 20 independent runs at every intersection. Table~\ref{tab:llm_parameter_sensitivity} reports test Fitness.}

\noindent
First, we test five LLM backends under identical settings. The results are backend-sensitive: GPT-4.1-mini and Gemini-2.5-Flash obtain the two best average ranks of 1.667 and 2.000, respectively.

\textcolor{red}{Second, we test the effects of the principle-informed prompt and external evaluator. The principles specify only general requirements for a reasonable delay formula; they do not provide its mathematical form or coefficient values. We compare four configurations: neither component, prompt only, evaluator only, and both components. Table~\ref{tab:llm_parameter_sensitivity} shows that using both produces the highest Fitness at all six intersections. Evaluator only also clearly outperforms prompt only, demonstrating the additional value of explicit verification beyond stating the principles in the prompt.}

\textcolor{red}{Third, we test temperatures of $0.0$, $0.3$, and $0.7$ with the principle instruction enabled or disabled. The instruction improves Fitness at every temperature. With the instruction enabled, no statistically significant differences are detected among the three temperatures.}

Finally, we test nine generation--population combinations. The $(20,20)$ and $(20,10)$ settings have the best average ranks, but neither significantly outperforms the default $(10,20)$. We retain $(10,20)$ for comparable Fitness with half the generations of $(20,20)$. \textcolor{red}{Under this default, a run averages $226.4\pm8.9$ attempted LLM API calls, $757\pm30.2$~s of cumulative API latency, and $456\pm30$~s of coefficient calibration (20 independent runs).}

\section{Conclusion}

Traffic delay estimation remains a fundamental yet challenging problem in intelligent transportation systems. This paper presented \textbf{CoSyDelay}, a cognitive symbolic regression framework for trustworthy delay modeling at signalized intersections. \textbf{CoSyDelay} evolves generalized symbolic expressions at the turning-movement level and calibrates them into physically consistent empirical models, which are then aggregated to support delay estimation at the approach and intersection levels. The framework combines domain-informed variable design, LLM-guided symbolic search, and a composite fitness function that evaluates both estimation accuracy and physical consistency. Experiments on six signalized intersections demonstrate that \textbf{CoSyDelay} achieves competitive delay estimation accuracy relative to analytical and data-driven baselines while producing physically consistent expressions. The ablation studies further support the use of turning-movement-level modeling, domain-informed features, and LLM-guided search. 

Future work will improve symbolic-search efficiency and extend \textbf{CoSyDelay} to larger-scale, real-time signal optimization. \textcolor{red}{In addition, geometric and queue information will be incorporated to better capture interactions between turning movements.}

\FloatBarrier
\bibliographystyle{IEEEtran}

\bibliography{References}

\end{document}